\documentclass[showpacs,twocolumn,superscriptaddress]{revtex4}

\usepackage{doi}
\usepackage{hyperref}
\hypersetup{
  colorlinks=true,        
  linkcolor=blue,         
  citecolor=cyan,         
}

\usepackage{graphicx}
\usepackage{dcolumn}
\usepackage{bm}
\usepackage{color}
\usepackage{enumitem}
\usepackage{amsmath}
\usepackage{amssymb}

\begin{document}
\title{Shadow and deflection angle of charged rotating black hole surrounded by perfect fluid dark matter}

\author{Farruh Atamurotov}
\email{atamurotov@yahoo.com,f.atamurotov@inha.uz}
\affiliation{Inha University in Tashkent, Ziyolilar 9, Tashkent 100170, Uzbekistan}
\affiliation{Akfa University, Kichik Halqa Yuli Street 17,  Tashkent 100095, Uzbekistan}
\affiliation{Department of Astronomy and Astrophysics, National University of Uzbekistan,Tashkent  100174, Uzbekistan}
\affiliation{Institute of Nuclear Physics, Tashkent 100214, Uzbekistan}
\author{Uma Papnoi} \email{uma.papnoi@gmail.com}
\affiliation{Gurukula Kangri (Deemed to be University), Haridwar 249404, Uttarakhand, India}
\affiliation{Kanoria PG Mahila Mahavidyalaya, Jaipur 302004, Rajasthan, India}
\author{Kimet Jusufi}
\email{kimet.jusufi@unite.edu.mk}
\affiliation{Physics Department, State University of Tetovo, Ilinden Street nn, 1200,
Tetovo, North Macedonia}
\begin{abstract}
We analysed the shadow cast by charged rotating black hole (BH) in presence of perfect fluid dark matter (PFDM). We studied the null geodesic equations and obtained the shadow of the charged rotating BH to see the effects of PFDM parameter $\gamma$, charge $Q$ and rotation parameter $a$, and it is noticed that the size as well as the shape of BH shadow is affected due to PFDM parameter, charge and rotation parameter. Thus, it is seen that the presence of dark matter around a BH affects its spacetime. We also investigated the influence of all the parameters (PFDM parameter $\gamma$, BHs charge $Q$ and rotational parameter $a$) on effective potential, energy emission by graphical representation, and compare all the results with the non rotating case in usual general relativity. To this end, we have also explored the effect of PFDM on the deflection angle and the size of Einstein rings.

\end{abstract}

\maketitle

\section{\bigskip INTRODUCTION}

General relativity (GR) is the geometric theory of gravitation which provides an unified description of gravity as a property of spacetime
\cite{Hartle03}. Black holes (BHs) are perhaps one of the most fascinating predictions of Einstein's theory of GR \cite{Chandrasekhar98a}. The BHs in GR and alternative theories of gravity exhibit the largest curvature of spacetime accessible to direct or indirect measurements. They are, therefore, ideal systems to test our theories of gravity under extreme
gravitational conditions. The existence of super massive BHs has been investigated extensively for nearly two decades, through
various esoteric astrophysical phenomena.  In 2019, an international collaboration of Event Horizon Telescope (EHT) has revealed the first shadow image of a
supermassive BH in the centre of the galaxy M87*  \cite{Akiyama19L1,Akiyama19L2,Akiyama19L3,Akiyama19L4,Akiyama19L5,Akiyama19L6} using very long baseline interferometer(VLBI) which shows a bright ring with a dark, central spot. Recently, the
second image was shown how an astrophysical environment,
like the magnetic field, distorts the image of the BH \cite{Akiyama19L12, Akiyama19L13}. Since these observations, in particular motivated by the Event Horizon Telescope (EHT) collaboration imaging of the shadow of M87* \cite{Akiyama19L1}, the investigation of BH shadows has become an active  field of research as the means to test GR and modified gravity. The study of phenomenon for BH shadow can provide the useful information about the various physical aspects of any given BH spacetime and the recent image from Event Horizon Telescope (EHT) \cite{Akiyama19L1,Akiyama19L2,Akiyama19L3,Akiyama19L4,Akiyama19L5,Akiyama19L6} marks a significant mile stone in the study of BH physics. The mathematical aspect of the very phenomenon which has been developed over the year \cite{Synge66a,Grenzebach14a,Papnoi14a,Stuchlik18a,Atamurotov16a,Abdujabbarov15a,Huang16a,Bisnovatyi18a,Moffat20a,Contreras20a,Chang20a,Dey20a} can be studied with observed data which can further help to hone the mathematical models of the extraordinary astronomical objects like a BH. This study therefore supports the earlier mathematical studies for existence of BHs in our universe as well as provides a possible way to verify one of the fundamental predictions of GR. We expect a good opportunity in near future to peek into the regime of strong gravity with ongoing observations like the Next Generation Very Large Array \cite{Hughes15a}, the Thirty Meter Telescope \cite{Sanders13a}, and the BlackHoleCam \cite{Goddi17a}.
EHT observation has triggered the study of gravitational lensing the BH shadow via probing the geometry of BHs.

Studying the BH shadow and gravitational lensing are most powerful astrophysical tools for investigating the strong field features of gravity which could provide a profound test of modified theories of
gravity in the strong field regimes. Gravitational deflection of light by rotating BHs has received significant attention due to the tremendous advancement of current observational facilities \cite{Virbha00a,Virbha02a,Vazquez04a,Perlick04a,Frittelli00a,Bozza01a,Bozza02b,Eiroa02b,Sotani15a,Babar21c,Kumar20a,Bisnovatyi10a,Atamurotov21a,Babar21d,Abdujabbarov17a,Gallego18a,Babar21a,Hakimov16a,Atamurotovjcap2021,Er14a,Rogers17a,Atamurotov21PFDM,Morozova13aa}.
The shape of BH shadows with various parameters in various theories of gravity have been investigated in \cite{Falcke00a,Bambi09a,Amarilla10a,Hioki09a,Abdujabbarov13a,Amarilla12a,Luminet79a,Amarilla13a,Abdujabbarov16a,Atamurotov13a,Tsukamoto18a,Atamurotov13c,Perlick18a,Wang18a,Cunha20a}.
It is very interesting to study the effects of
plasma background in BH shadows. In recent years the study of astrophysical processes in plasma medium surrounding BH has become very interesting and important due to the evidence for the presence of BHs at the centers of galaxies. Observing the BH shadow in inhomogeneous and homogeneous plasma around BHs has been recently studied in as extension of vacuum studies (see, e.g. \cite{Perlick15a, Perlick17a,Chowdhuri20a,Atamurotov15a,Babar20a,Fathi21a,Abdujabbarov17b,Atamurotov21Axion}). The influence of a plasma on the shadow of a spherically symmetric BH is investigated by Perlick et al. \cite{Perlick15a}.

BHs can be characterized by at most three parameters, i.e., BH mass $M$, rotation parameter $a$ and electric charge $Q$ astrophysically \cite{Zakharov2018,Zakharov21} and the first two parameters are constrained by the BH mass. Astrophysical BHs can be regarded as highly spinning which has been seen in recent astrophysical observations (see for example \cite{Walton13,Patrick11,Bambi2017}). Various astrophysical mechanisms shows that BHs possess charge. Positive net electric charge in BHs is due to the balance between Coulomb and gravitational forces for charged particles near the surface of the compact object \cite{bally78} and due to the matter, which gets charged as that of the irradiating photons \cite{Weigartner2006}. Shadow of BH in presence of charge was first studied in \cite{deviers}, later, Zakharov  obtained some constraint on the BH charge using the EHT observational data in \cite{Zakharov2018,Zakharov21}.

Further, the cosmological evidences shows that the universe comprised of 68\% dark energy \cite{Hunterer18a}, 27\% dark matter \cite{Bertone18a}. It is doubtful to think that dark energy could affect the surrounding of a BH, but dark matter could produce an effective fluid around the compact object. Dark matter field can exist in the environment of supermassive BHs. Though dark matter has not been directly detected, observations has provided strong evidence that dark matter can exist in the environment of giant elliptical and spiral galaxies \cite{23}. Based on the theoretical analysis and astrophysical data, it is believed that dark matter contributes to approximately up to $90 \%$ mass of the galaxy, while the rest is the luminous matter composed of baryonic matter \cite{24}. Astrophysical observations indicate that giant elliptical and spiral galaxies are embedded in a giant dark matter halo \cite{Akiyama19L1,Akiyama19L6}. Several BH solutions involved the dark matter contribution in the background geometry has been studied. The modified BH metric  for the Schwarzschild BH in quintessence field has been obtained \cite{Kiselev03b} and its Quasinormal modes in \cite{Zhang06a}.  Later, the circular geodesic orbits and effect of external magnetic field on particle acceleration around quintessential rotating BHs has been discussed in  \cite{Toshmatov17a,Shaymatov18a}. It is assumed that dark matter is present around BHs as halos, for these case BH shadows has also been calculated \cite{Jusufi19a,Konoplya19a}. The cold dark matter around BH in phantom field background have been obtained in \cite{Li12a}. Recently perfect fluid model \cite{Kiselev03a} has been studied to see the impact of the dark matter on various observables related to the BHs. Shadow cast by rotating BHs in perfect fluid dark matter (PFDM) with and without cosmological constant along with non-rotating charged BHs has been studied in \cite{Haroon19a,Wang18a} and circular geodesic and energy conditions in a rotating charged BH in PFDM are also deeply investigated by Das et al. \cite{Das21a}. The geodesics around BH in PFDM can led to interesting observations for various BH spacetime in usual GR and modified theories. It is well known that in past two three years several results have been observed by VLBI under EHT after observing the first image of M87* candidate black hole. From these observational results black hole spin can be easily explained, but the information regarding the other parameters of black hole are yet to be explained, right now we can analyse only via theoretical concept. Hopefully in near future observational results will be obtained to study the other parameters as well. Under this idea there are  several scientists working with modified theory and getting constrain to compare with observational data. From our theoretical results also we can get more insights on black hole parameters using radius and distortion parameters of black hole. Consequently this will help us to explain the nature of the dark matter and charged black hole.
We have considered the charged BH in PFDM and obtained its rotating counterpart using \cite{Das21a, Newman65a} to investigate the optical properties by studying the photon motion. By using optical properties we may get a constraint on charge $Q$, PFDM parameter $\gamma$ and rotation parameter $a$ with more accurate measurements in the near future observational data.

The paper is organized as follows: In Sec. \ref{introductionsec}, we have discussed the
charged rotating BH solution in perfect fluid dark matter. In Sec. \ref{Shadowsec}, we have presented the particle motion around the BH solution to discuss the shadow of the BH. The Emission energy from the BH is investigated in Sec. \ref{emissionenergysec}. In Sec. \ref{deflectionsec} we study the deflection of particles and light. In Sec. \ref{ringsec}, we consider the effect of PFDM on the Einstein rings. Finally, we have concluded with the obtained results in  Sec~\ref{conclusionsec}.
 We have used units that fix the speed of light and the gravitational constant via $8\pi G = c^4 = 1$.

\section{Charged BH in perfect fluid dark matter}\label{introductionsec}
The action for the gravity theory minimally coupled with gauge field in  PFDM reads as \cite{Kiselev03b,Kiselev03a,Li12a}
\begin{eqnarray}
\mathcal{I}=\int dx^4\sqrt{-g}\left( \frac{1}{16\pi G} R+\frac{1}{4} F^{\mu \nu}F_{\mu \nu} + \mathcal{L}_{DM}\right).
\end{eqnarray}
Here, $g$ = det($g_{ab}$) is the determinant of the metric tensor, $R$ is the Ricci scalar, $G$ is the Newton's gravitational constant, $F_{\mu \nu}= \partial_\mu A_\nu -\partial_\nu A_\mu$ ($A_\mu$ is the gauge potential) is the Maxwell field strength and $\mathcal{L}_{DM}$ is the Lagrangian density for PFDM. Using the variation fro action principle, Einstein field equations are obtained as
\begin{eqnarray}
R_{\mu \nu}-\frac{1}{2}g_{\mu \nu}R = - 8\pi G (T_{\mu \nu}^M +T_{\mu \nu}^{DM}), \nonumber \\
= - 8 \pi G T_{\mu \nu},
\end{eqnarray}
\begin{eqnarray}
F^{\mu \nu}_{\nu}=0, \nonumber \\
F^{\mu \nu ;\alpha} +F^{\nu \alpha ;\mu}+F^{\alpha \mu ; \nu} =0.
\end{eqnarray} Here, $T_{\mu \nu}^M$ is the energy-momentum tensor for ordinary matter and $T_{\mu \nu}^{DM}$ is the energy-momentum tensor for PFDM \cite{Zhaoyi18a} given as
 \begin{eqnarray}
T^\mu_\nu =g^{\mu \nu} T_{\mu \nu}, \nonumber \\
T^t_t = -\rho, \;\;\; T^r_r = T^{\theta}_{\theta} = T^{\phi}_{\phi} = P.
\end{eqnarray} Further, following \cite{Li12a,Zhaoyi18a} we assume that $T^r_r=T^{\theta}_{\theta} = T^{\phi}_{\phi} = T^t_t (1-\delta)$ where $\delta$ is a constant. For the Reissner-Nordstr$\Ddot{o}$m metric in
PFDM is
\begin{eqnarray}\label{pfdm1}
ds^2 = -f(r) dt^2 + f^{-1}(r) dr^2 +r^2 (d\theta^2 +\sin^2\theta d\phi^2),
\end{eqnarray} with
\begin{eqnarray}
f(r) = 1-\frac{2 M}{r} +\frac{Q^2}{r^2} + \frac{\gamma}{r} \ln(\frac{r}{\gamma}),
\end{eqnarray}
where $M$ is the mass of the BH, $Q$ is the charge of the BH and $\gamma$ is the PFDM parameter.
\subsection{Generating mechanism for charged rotating black hole}

Considering Reissner-Nordst$\Ddot{o}$rm as the seed solution and applying Newman and Janis \cite{Newman65a} to construct its rotating counterpart, first step is to write the metric (\ref{pfdm1}) in advanced Eddington-Finkelstein coordinates using the coordinate transformation:
\begin{eqnarray}
du &=& dt- f(r)^{-1} dr, \label{eq:du}
\end{eqnarray}
which gives
\begin{eqnarray}
ds^2 = f(r) du^2+ 2 du dr  -r^2(d\theta^2 + \sin^2 \theta d\phi^2). \label{eq:me3}
\end{eqnarray}
The metric (\ref{eq:me3}) can be
written in terms of a null tetrad $ Z^a =
(l^{\mu},\;n^{\mu},\;m^{\mu},\;\bar{m}^{\mu}), $ \cite{Newman65a} as
\begin{eqnarray}
{g}^{\mu \nu} = l^{\mu} n^{\nu} + l^{\nu} n^{\mu} - m^{\mu} \bar{m}^{\nu},
\label{NPmetric}
\end{eqnarray}

 where null tetrad are \begin{eqnarray*}
l^{\mu} &=& \delta^{\mu}_r,\\
n^{\mu} &=& \left[ \delta^{\mu}_u -  1-\frac{2 M}{r} +\frac{Q^2}{r^2} + \frac{\gamma}{r} \ln(\frac{r}{\gamma}) \delta^{\mu}_r \right ],\\
 m^{\mu} &=& \frac{1}{\sqrt{2}r} \left( \delta^{\mu}_{\theta}
  + \frac{i}{\sin \theta} \delta^{\mu}_{\theta} \right).\\
\end{eqnarray*}
Note
$l^{\mu}$ and $n^{\mu}$ are real, $m^{\mu}$, $\bar{m}^{\mu}$ are mutual complex
conjugate. This tetrad is orthonormal and obeying metric
conditions.
\begin{eqnarray}
l_{\mu}l^{\mu} = n_{\mu}n^{\mu} = ({m})_{\mu} ({m})^{\mu} = (\bar{m})_{\mu} (\bar{m})^{\mu}= 0,  \nonumber \\
l_{\mu} n^{\mu} = 1, \; ({m})_{\mu} (\bar{m})^{\mu} = 1. \nonumber
\end{eqnarray}
Now we allow for some $r$ factor in the null vectors to take on complex values. Following \cite{Newman65b}, we rewrite the null vectors in the form
\begin{eqnarray*}
l^{\mu} &=& \delta^{\mu}_r, \\
n^{\mu} &=& \left[ \delta^{\mu}_u -  1- M (\frac{1}{r}+\frac{1}{\bar{r}}) +\frac{Q^2}{r \bar{r}} + \gamma (\frac{1}{r}+\frac{1}{\bar{r}}) \ln(\frac{r}{\gamma}) \delta^{\mu}_r \right ], \\
m^{\mu} &=& \frac{1}{\sqrt{2}r} \left( \delta^{\mu}_{\theta}
  + \frac{i}{\sin\theta} \delta^{\mu}_{\theta} \right),\\
\end{eqnarray*}
with $\bar{r}$ is the complex conjugate of $r$. Next defining a new set of coordinates $(u',r',\theta')$, to perform the complex coordinate  transformation, using
the relations

\begin{equation}\label{transf}
u' = u - ia\cos\theta, \\
r' = r + ia\cos\theta, \\
\theta' = \theta,
\phi ' = \phi.
\end{equation}
Simultaneously let null tetrad vectors $Z^{\mu}$ undergo a transformation $Z^{\mu} = Z'^{\mu}{\partial x'^{\mu}}/{\partial x^{\nu}} $ in the usual way, which gives

\begin{eqnarray*}
l^{\mu} &=& \delta^{\mu}_r, \\
n^{\mu} &=&  \left[ \delta^{\mu}_u -  1-\frac{2 M}{\rho^2} +\frac{Q^2}{\rho^2} + \frac{\gamma}{r} \ln(\frac{r}{\gamma}) \delta^{\mu}_r \right ],\\
 m^{\mu} &=& \frac{1}{\sqrt{2}(r+ia\cos\theta)} \left(ia(\delta^{\mu}_u-\delta^{\mu}_r)\sin\theta + \delta^a_{\theta} + \frac{i}{\sin\theta} \delta^{\mu}_{\theta} \right),\\
 \end{eqnarray*}
where we have drop the primes. A new metric is discovered from the new null tetrad  in Kerr like coordinates \cite{X88a}.  A
further simplification is made by using another coordinate transformation as in Ref. \cite{X88a} for the simplification of the metric which leaves only one off-diagonal element and the metric obtained for charged rotating BH in PFDM is: \cite{Das21a}
\begin{eqnarray}
ds^2 &=& -\frac{1}{\rho^2}\Big(\Delta -a^2 \sin^2\theta \Big)dt^2+ \frac{\rho^2}{\Delta} dr^2+\rho^2 d\theta^2\nonumber
\\ &&-\frac{2a \sin^2\theta}{\rho^2}\Big[2 M r -Q^2 -\gamma r \ln(\frac{r}{\gamma})\Big]dt d\phi  \nonumber \\ && +\sin^2\theta \Big[r^2 +a^2 +\frac{a^2 \sin^2\theta}{\rho^2}\Big(2 M r -Q^2 \nonumber \\ &&-\gamma r \ln(\frac{r}{\gamma}) \Big)\Big] d\phi^2.
\end{eqnarray}
Here $a$ is the rotational parameter with $\Delta$ and $\rho$ given by
\begin{eqnarray}
\Delta & =& r^2 +a^2 -2M r +Q^2 + \gamma r \ln (\frac{r}{\gamma}),\nonumber \\
\;\;\; \rho^2 &=& r^2 +a^2 \sin^2\theta.
\end{eqnarray}

The BH horizon is determined by taking roots of the equation $\Delta=0$ as

\begin{eqnarray}\label{1}
r^2 +a^2 -2M r +Q^2 = \gamma r \ln (\frac{r}{\gamma})
\end{eqnarray}
which corresponds to the two horizons, the inner horizon $r_-$ and the outer horizon $r_+$. It can be noticed that the size of the two horizons depends on $\gamma$. Since $\gamma$ does not produce new horizons, above equation has two roots and one extreme value point. Now, taking the derivative of Eq (\ref{1}) with respect to $r$
results in \cite{Xu2018}
\begin{equation}
    r-M = \frac{\gamma}{2}ln(\frac{r}{\gamma})+\frac{\gamma}{2}.
\end{equation}
For $\gamma >0$, the maximum of $\gamma$ satisfies \cite{Xu2018}
\begin{equation}
\frac{2}{\gamma_{max}}(r-M)-1 = ln(\frac{2M}{\gamma_{max}})
\end{equation}

\begin{figure*}
 \begin{center}
   \includegraphics[scale=0.6]{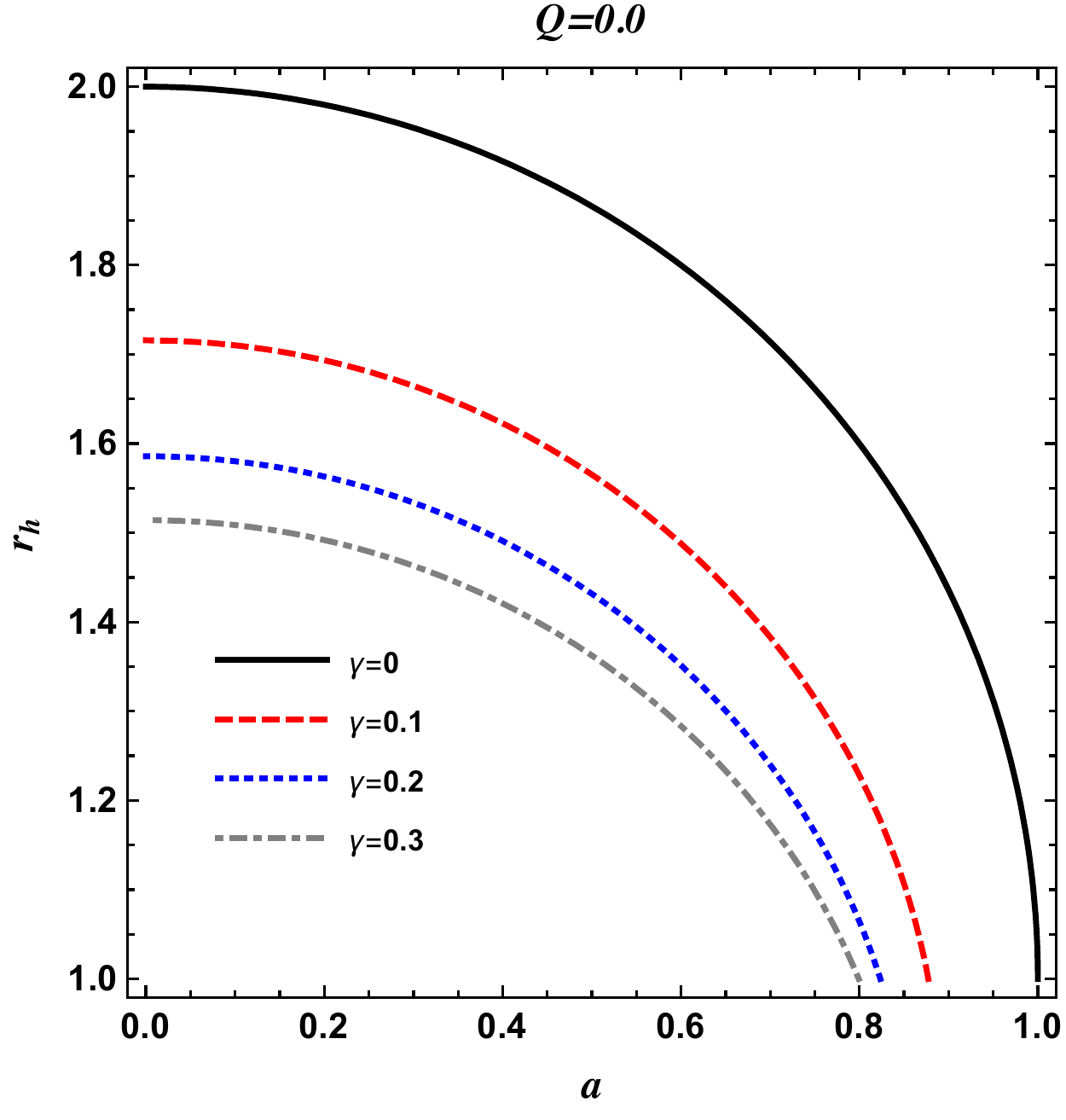}
   \includegraphics[scale=0.6]{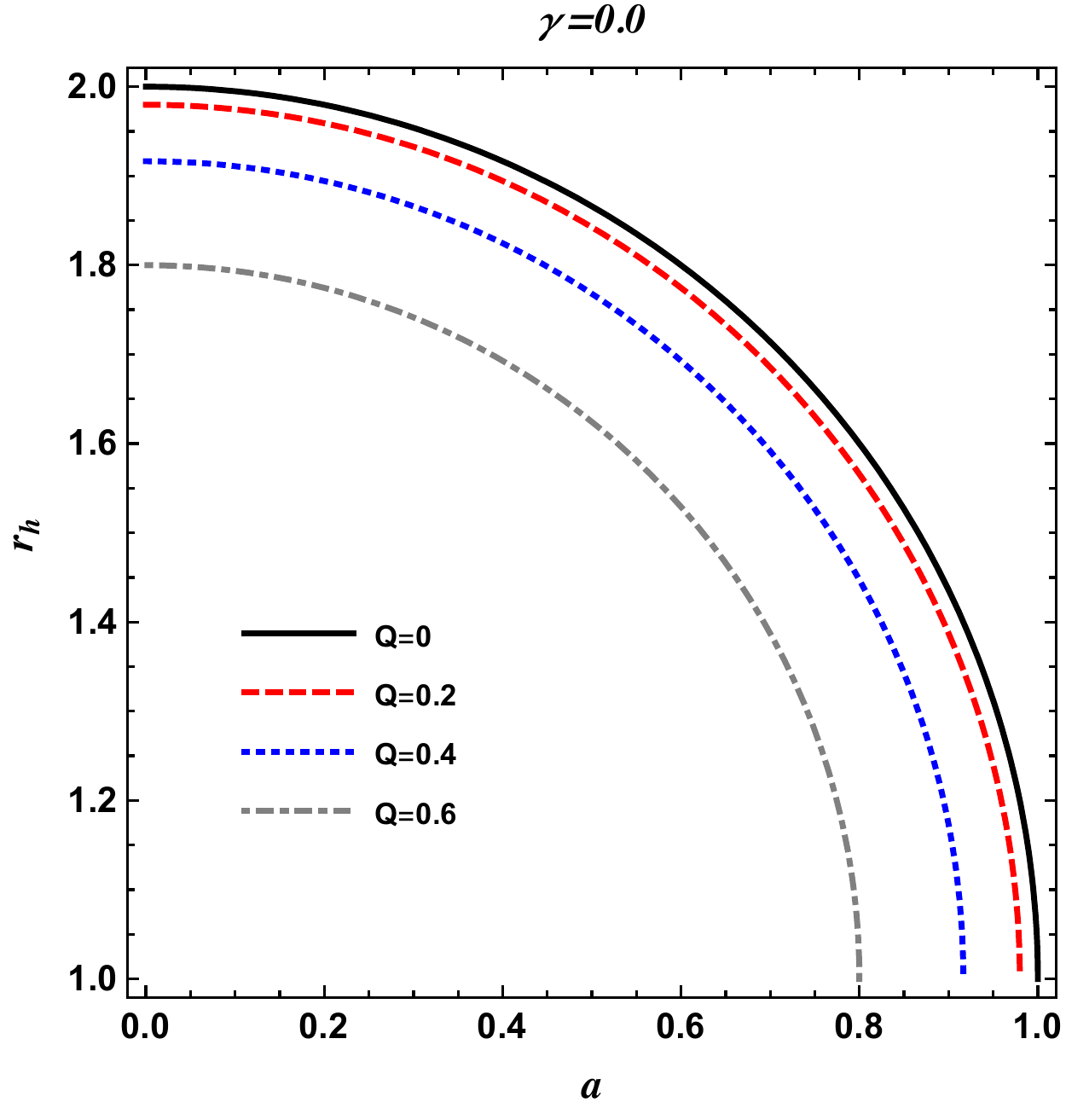}

   \includegraphics[scale=0.6]{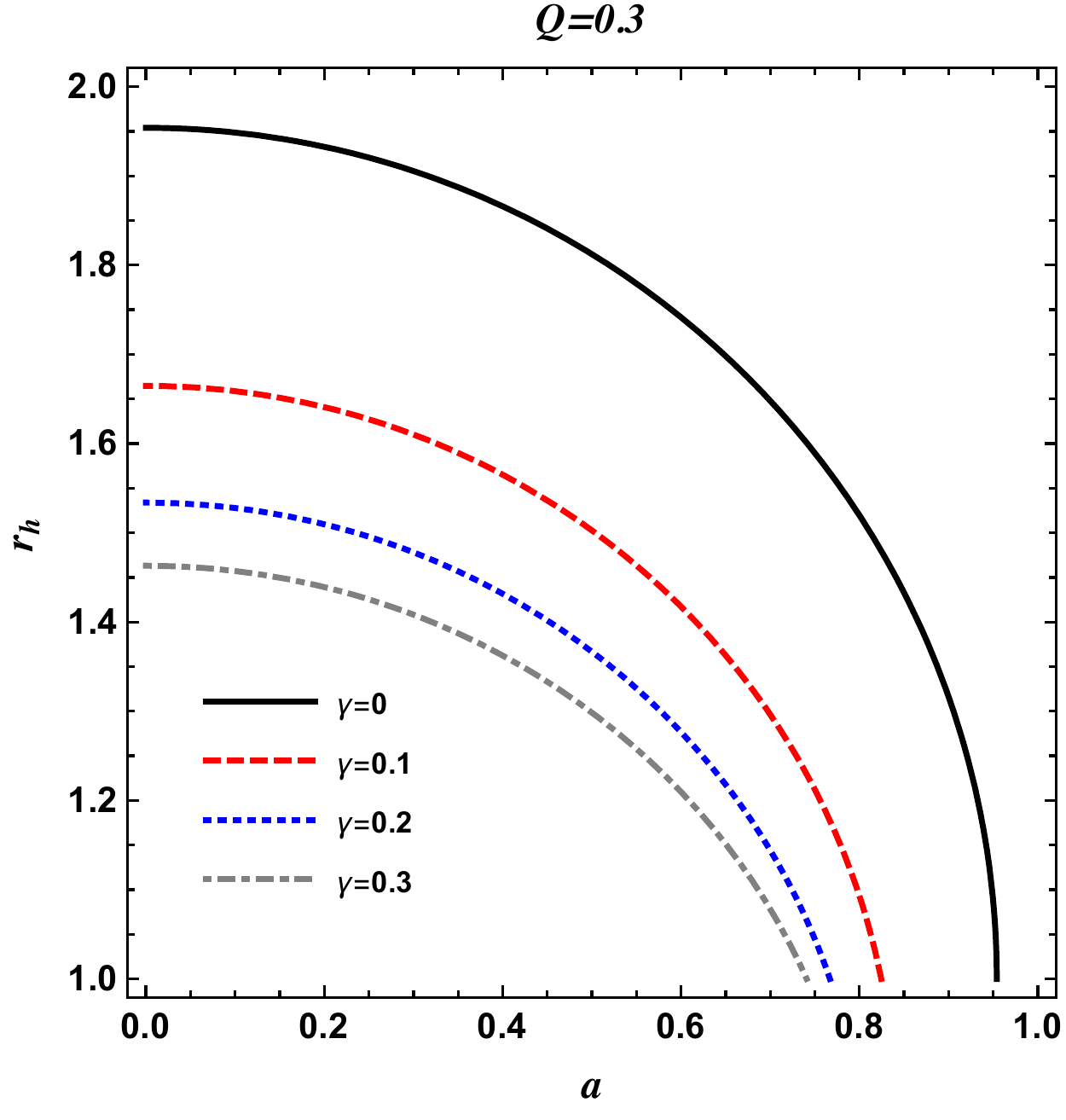}
   \includegraphics[scale=0.6]{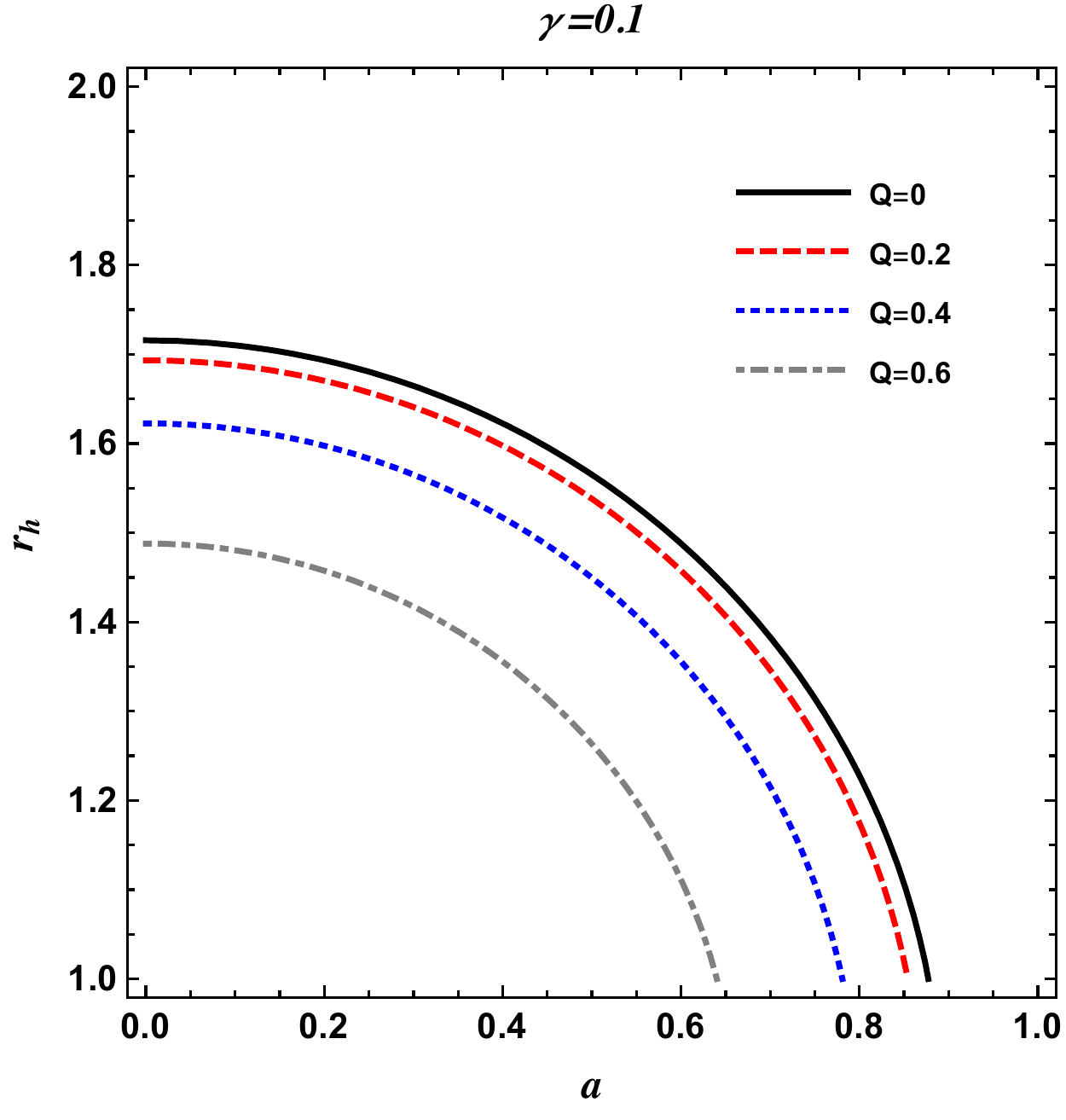}
  \end{center}
\caption{Radial dependence of the rotation parameter for different values of PFDM parameter at fixed $Q$ (Both left panels) and for different values of $Q$ at fixed $\gamma$ (Both right panels). }\label{horizon}
\end{figure*}

In fig. (\ref{horizon}), we have shown the dependence of spin parameter $a$ on the radial coordinate $r$ for the different values of charge $Q$ and PFDM parameter $\gamma$. It can be seen that both charge and PFDM parameter shows an influence on the horizon radius of the BH.

\subsection{Geodesics and photon orbits}
To investigate the shape of shadow, we begin with analysing the general orbit of photon around a BH. We study the separability of the Hamilton-Jacobi equation as indicated by Carter in \cite{Carter68a}. It is well known that the
Hamilton-Jacobi equation with the metric tensor $g_{\mu\nu}$ takes the
following general form:
\begin{eqnarray}\label{Hamilton-Jacobieq}
-\frac{\partial S}{\partial \lambda} =
\frac{1}{2} g^{\mu \nu} \frac{\partial S}{\partial x^\mu}
\frac{\partial S}{\partial x^\nu},
\end{eqnarray}
with $\lambda$ as an affine parameter and an action $S$ which can
be written as:
\begin{eqnarray}\label{Hamilton-Jacobiac}
S= \frac{1}{2} m^2 \lambda -Et+ L\phi  +
S_\theta (\theta) + S_r (r) \ ,
\end{eqnarray}
in order to separate the equation of motion. Here, $E$ and $L$, which corresponds to the specific energy and angular momentum. These are constants of motion related to the associated Killing vectors $\partial/\partial t$ and $\partial/\partial \phi$.

For the photon the mass of the particle $m$ is equal to zero
($m=0$). Since we need to study the shadow of the charged rotating BH in PFDM, therefore solving the Hamilton Jacobi equation for null geodesics,  (\ref{Hamilton-Jacobieq}) and
(\ref{Hamilton-Jacobiac}), the functions $R$ and $\Theta$ corresponding to radial and $\theta$ motion are obtained as



\begin{eqnarray}\label{R}
R=[(r^2+a^2)E - a L]^2-\Delta[{\cal K}+(a E-L)^2]
\end{eqnarray}
and
\begin{eqnarray}
\Theta =(a^2 E^2 - \frac{L^2}{\sin^2\theta })\cos^2\theta + {\cal K}.
\end{eqnarray}
Here ${\cal K}$ is the constant of  separation termed as Carter's constant.
\begin{figure*}
 \begin{center}
   \includegraphics[scale=0.45]{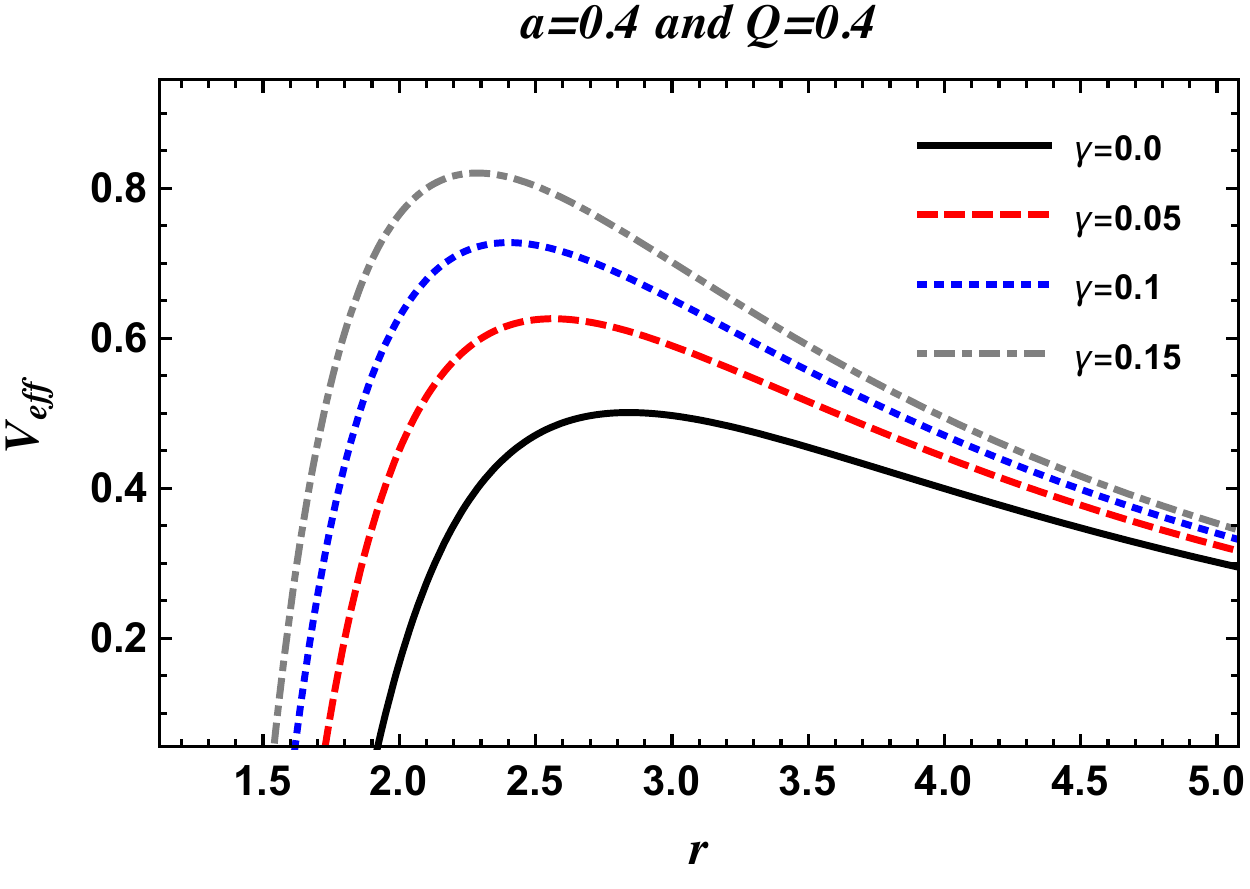}
   \includegraphics[scale=0.45]{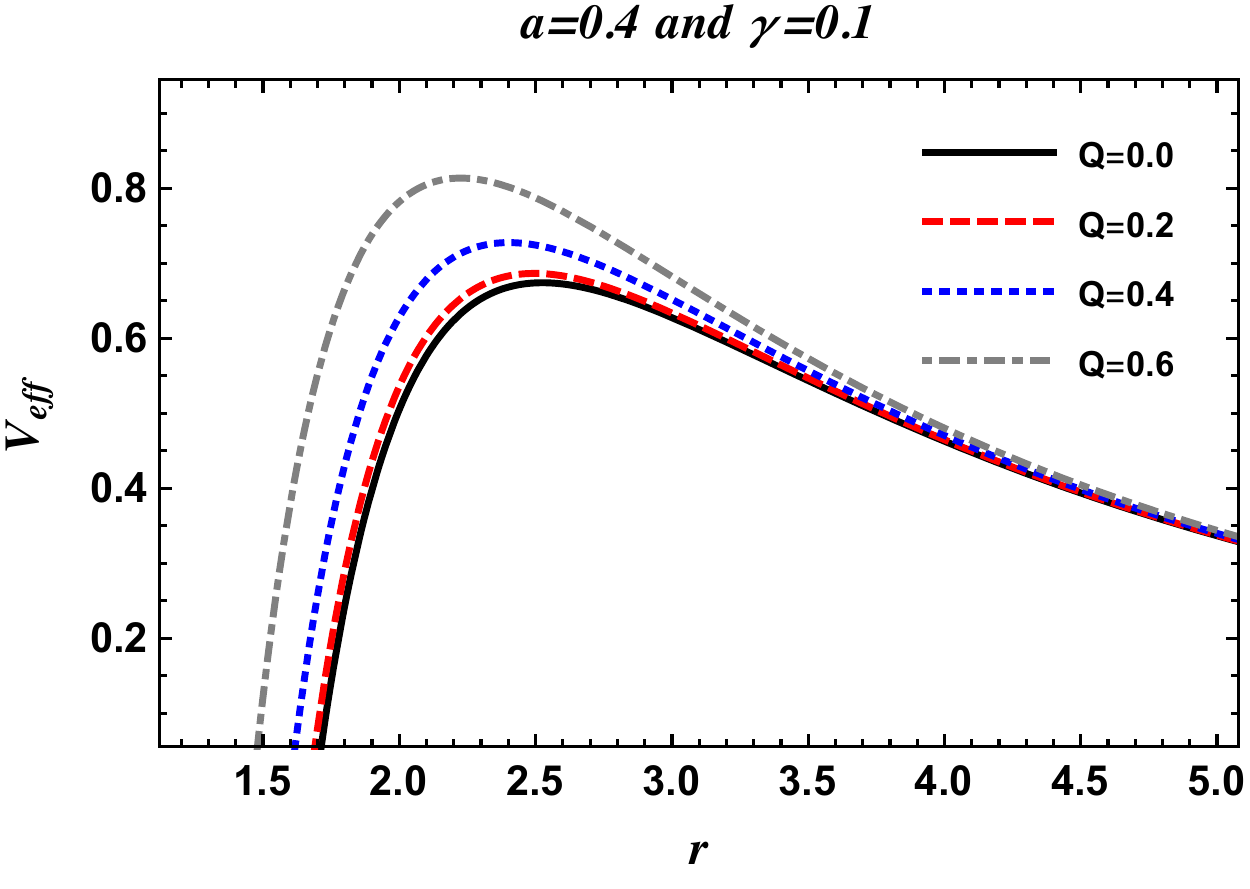}
   \includegraphics[scale=0.45]{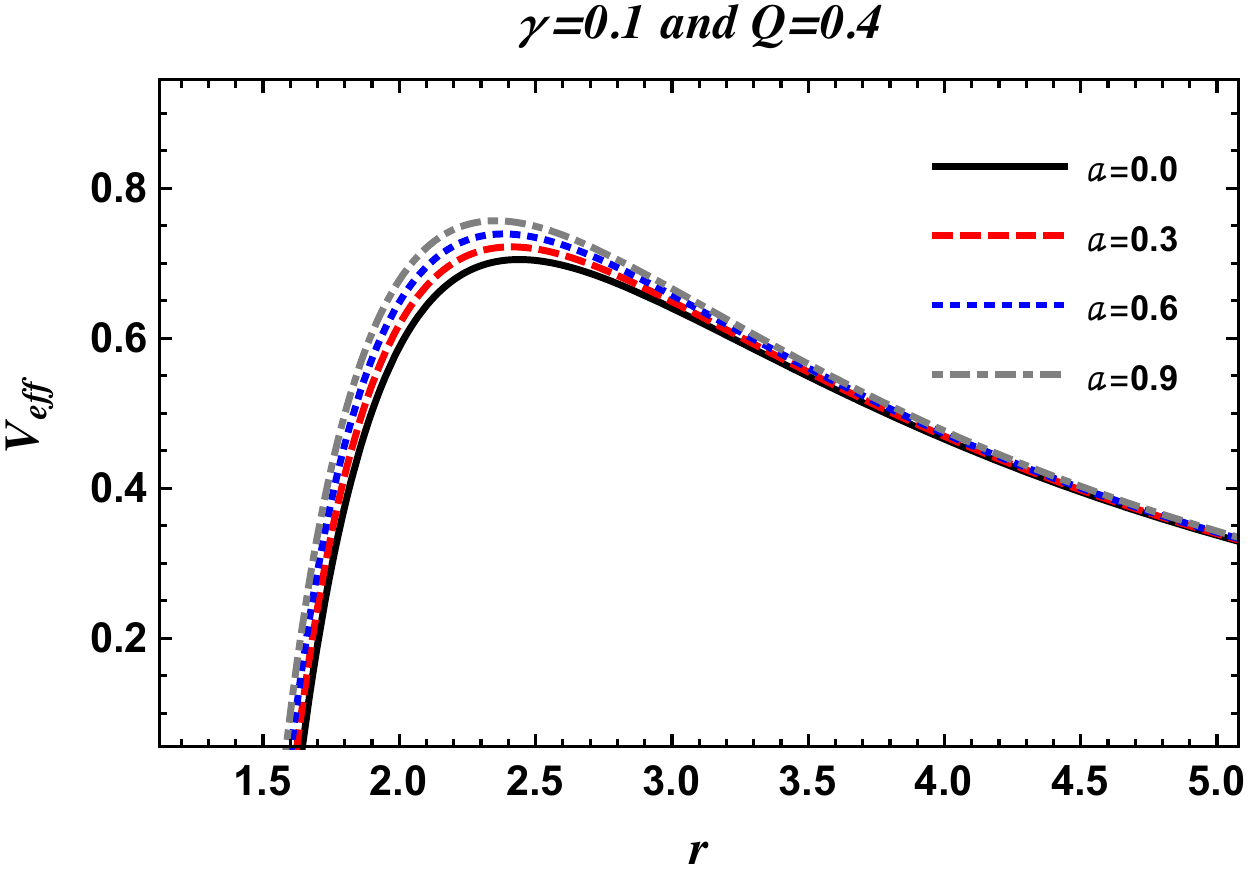}
   \end{center}
\caption{Plots showing the radial dependence of effective  potential for different value of $\gamma$ with $a=0.4, Q=0.4$ (Left panel), for different values of $Q$ with $a=0.4, \gamma =0.1$ (middle panel) and for different value of $a$ at $Q=0.4, \gamma =0.1$(Right panel).}\label{effpotential}
\end{figure*}

The geodesic equations obtained are:
\begin{eqnarray}\label{EOM}
\rho^4 \dot{\theta}^2 &=& \Theta, \nonumber \\
\rho^4 \dot{r}^2 &=& R  ,\nonumber \\
\rho^2 \dot{t} &=& \frac{r^2+a^2}{\Delta}\Big[E(r^2+a^2)-a L\Big]+a(L-a E \sin^2\theta) ,\nonumber \\
\rho^2 \dot{\phi} &=&  \frac{a}{\Delta}\Big[E(r^2+a^2)-a L\Big]+(\frac{L}{\sin^2\theta}-a E).
\end{eqnarray}
In order to analyse the shadow of the BH, we need to study the radial motion of photons around the BH. Thus rewriting the radial equation as\begin{eqnarray}
\rho^2 \dot{r} &=& \sqrt{R}.
\end{eqnarray}
The effective potential ($V_{eff}=-\dot{r}^2/2$) for equatorial plane ($\theta=\pi/2$) reads \begin{eqnarray}\label{veff}
V_{eff}&=&\frac{\Delta ({\cal K} + (L - a E)^2)-((a^2 + r^2) E - a L)^2 }{2r^4}
\end{eqnarray}

The general behavior of the effective potential for charged rotating BH  as a function of $r$ for different
values of PFDM parameter, charge and the rotation parameter is shown in Fig. \ref{effpotential}. It is worth to note that for all the case the effective potential shows a maxima which correspond to the unstable circular orbits. It is seen
that with the increase in the value of PFDM parameter $\gamma$, charge $Q$ and rotation parameter $a$ the peak of the graph is shifting towards the left, which signifies that
the circular orbits are shifting towards the central object.

Since the constants
of motion $E$, and $L$ and the constant of
separability ${\cal K}$ play the role of the characteristics of
each photon orbit, so one can define the
 impact parameters $\xi=L/E$,
 and $\eta={\cal K}/E^2$ for photon orbits.

It is well known that the unstable photon orbits are responsible for the shadow cast by a BH and hence the boundary of the shadow attained by charged rotating BH can be derived using the conditions
\begin{equation}\label{condition}
V_{eff}=0, \;\;\; \partial V_{eff}/\partial r=0 \nonumber \\ (\textrm{or}
R(r)=0=\partial R(r)/\partial r)\ .
\end{equation}

Using the Eq.~(\ref{R}) in Eq. (\ref{condition}), one can obtain the contour of the shadow as
\begin{widetext}
\begin{eqnarray}
\xi=\frac{-2 a^2 M-2 a^2 r-4 Q^2 r+6 M r^2-2 r^3+a^2 \gamma +r^2 \gamma +a^2 \gamma  \ln\left[\frac{r}{\gamma }\right]-3 r^2 \gamma  \ln\left[\frac{r}{\gamma }\right]}{-2 a M+2 a r+a \gamma +a \gamma  \ln\left[\frac{r}{\gamma }\right]},
\end{eqnarray}
\begin{eqnarray}
\eta &=& \frac{1}{\Big(-2 a^2 M+2 a^2 r+a^2 \gamma +a^2 \gamma  \ln\Big[\frac{r}{\gamma }\Big]\Big)}\Big(-4 a^2 Q^2 r+4 a^2 M r^2-4 Q^2 r^3+6 M r^4-2 r^5+2 a^2 r^2 \gamma \nonumber \\ && +r^4 \gamma -2 a^2 r^2 \gamma  \ln\Big[\frac{r}{\gamma }\Big]-3 r^4 \gamma  \ln\Big[\frac{r}{\gamma }\Big] \nonumber \\ && +\frac{4 a Q^2 r \Big(-2 a^2 M-2 a^2 r-4 Q^2 r+6 M r^2-2 r^3+a^2 \gamma +r^2 \gamma +a^2 \gamma  \ln\Big[\frac{r}{\gamma }\Big]-3 r^2 \gamma  \ln\Big[\frac{r}{\gamma }\Big]\Big)}{-2 a M+2 a r+a \gamma +a \gamma  \ln\Big[\frac{r}{\gamma }\Big]} \nonumber \\ && -\frac{4 a M r^2 \Big(-2 a^2 M-2 a^2 r-4 Q^2 r+6 M r^2-2 r^3+a^2 \gamma +r^2 \gamma +a^2 \gamma  \ln\Big[\frac{r}{\gamma }\Big]-3 r^2 \gamma  \ln\Big[\frac{r}{\gamma }\Big]\Big)}{-2 a M+2 a r+a \gamma +a \gamma  \ln\Big[\frac{r}{\gamma }\Big]} \nonumber \\ && -\frac{2 a r^2 \gamma  \Big(-2 a^2 M-2 a^2 r-4 Q^2 r+6 M r^2-2 r^3+a^2 \gamma +r^2 \gamma +a^2 \gamma  \ln\Big[\frac{r}{\gamma }\Big]-3 r^2 \gamma  \ln\Big[\frac{r}{\gamma }\Big]\Big)}{-2 a M+2 a r+a \gamma +a \gamma  \ln\Big[\frac{r}{\gamma }\Big]} \nonumber \\ && +\frac{2 a r^2 \gamma  \ln\Big[\frac{r}{\gamma }\Big] \Big(-2 a^2 M-2 a^2 r-4 Q^2 r+6 M r^2-2 r^3+a^2 \gamma +r^2 \gamma +a^2 \gamma  \ln\Big[\frac{r}{\gamma }\Big]-3 r^2 \gamma  \ln\Big[\frac{r}{\gamma }\Big]\Big)}{-2 a M+2 a r+a \gamma +a \gamma  \ln\Big[\frac{r}{\gamma }\Big]}\Big).
\end{eqnarray}
\end{widetext}

\begin{figure*}
 \begin{center}
   \includegraphics[scale=0.5]{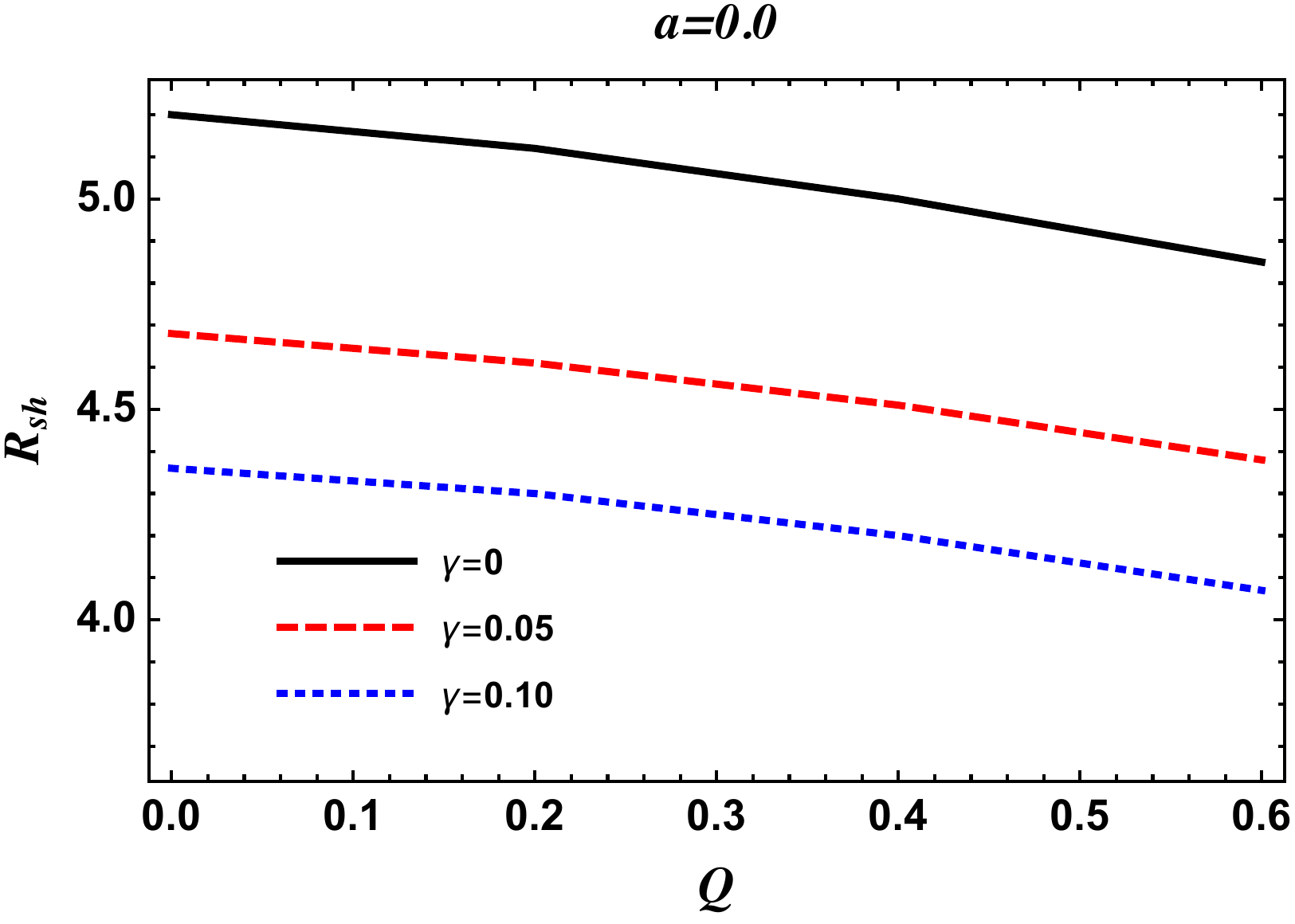}
   \includegraphics[scale=0.5]{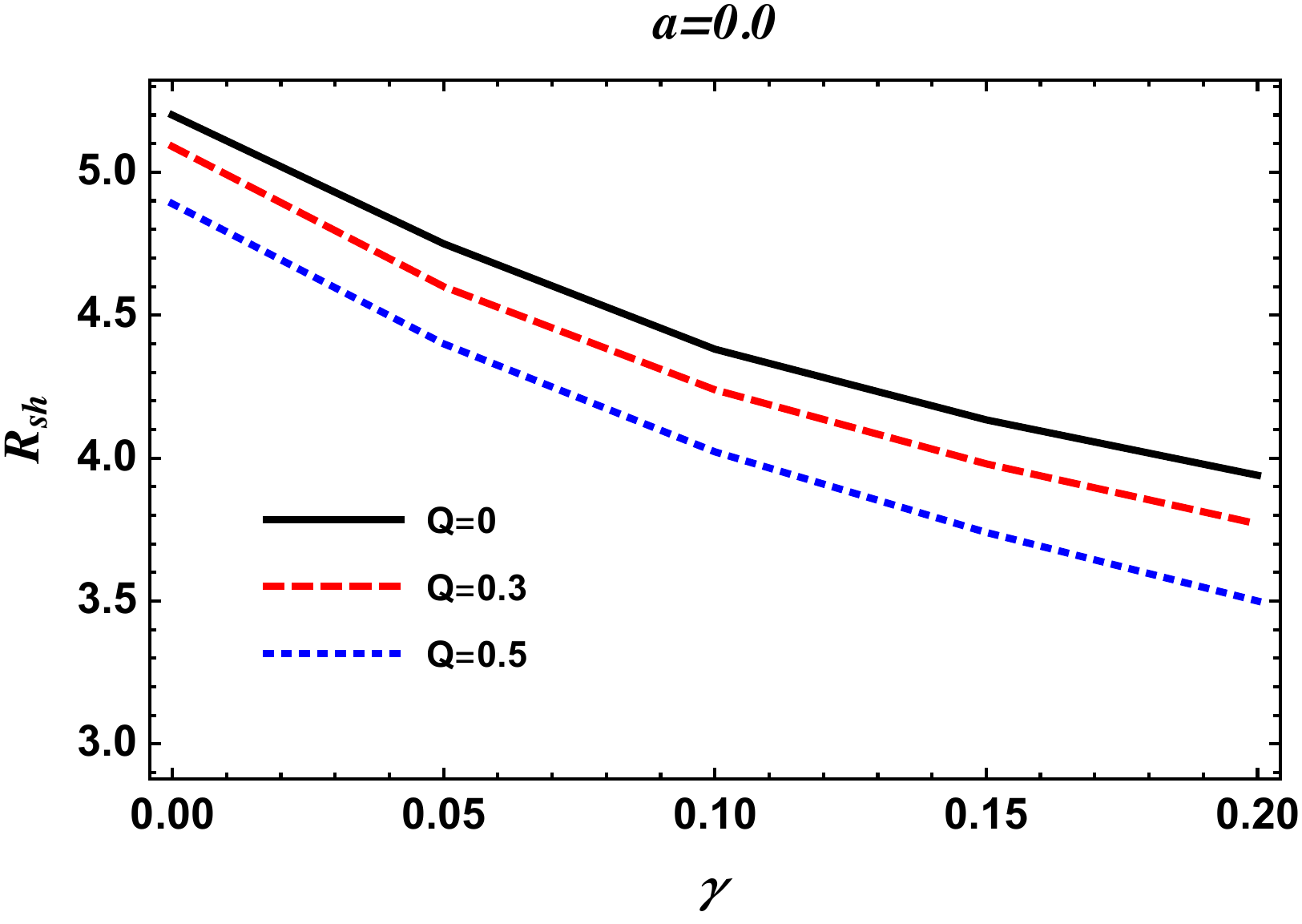}
  \end{center}
\caption{Plots showing the variation of shadow for non rotating charged BH with $Q$ at $\gamma = 0; 0.05; 0.1$ (Left panel), with $\gamma$ at $0; 0.3; 0.5$ (Right panel).}\label{shadowa0}
\end{figure*}

\begin{figure*}
 \begin{center}
   \includegraphics[scale=0.45]{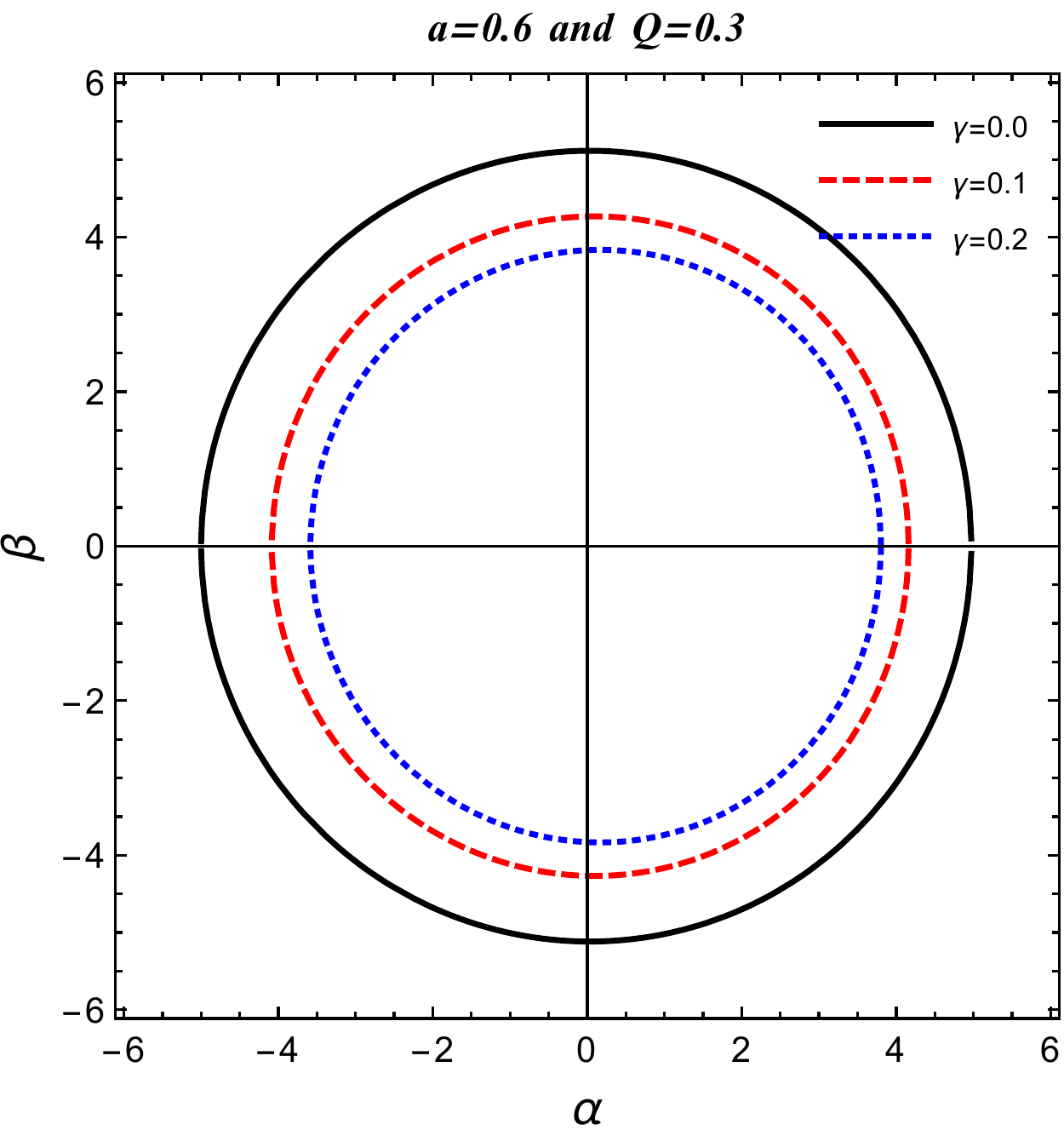}
   \includegraphics[scale=0.45]{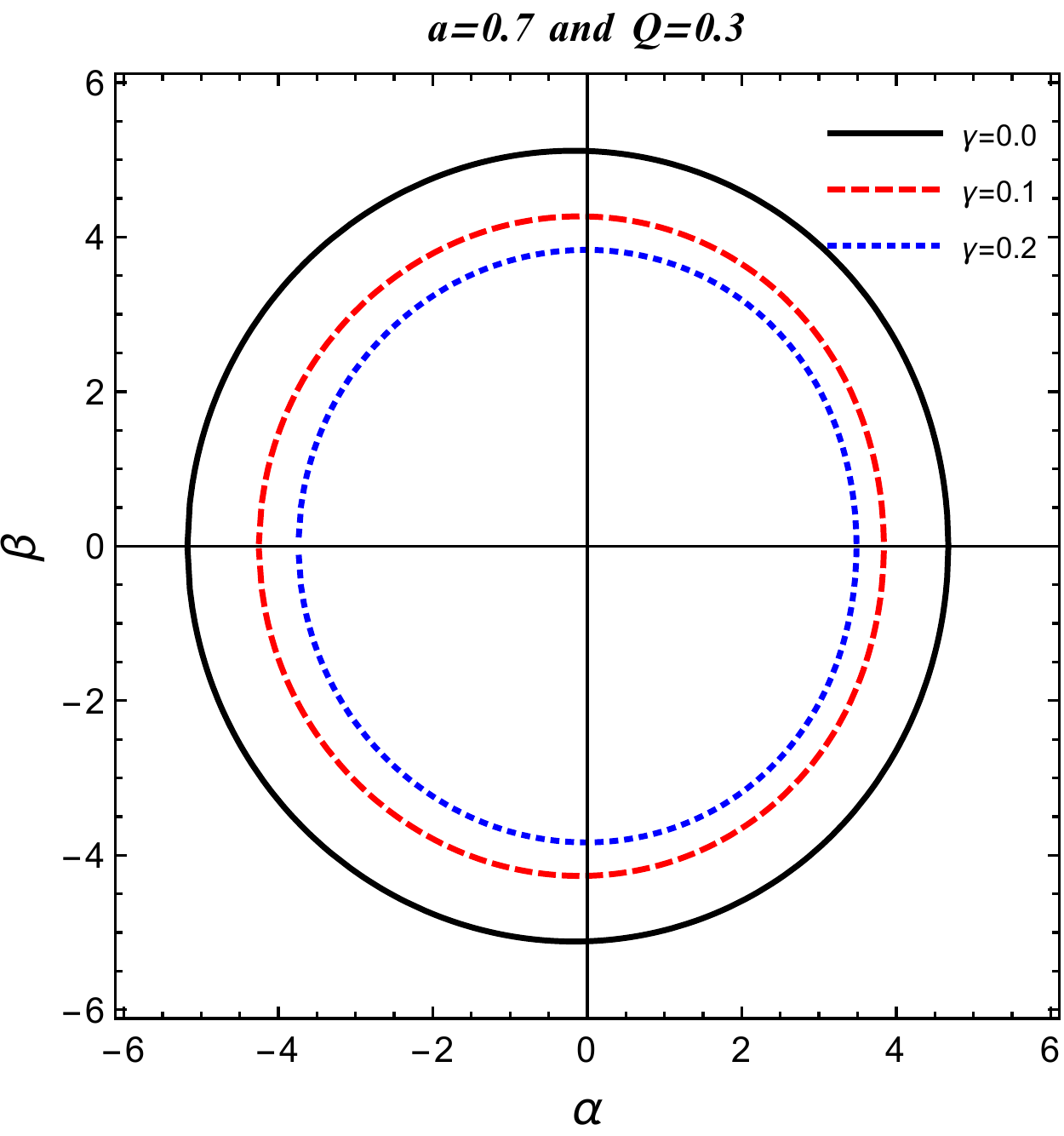}
    \includegraphics[scale=0.45]{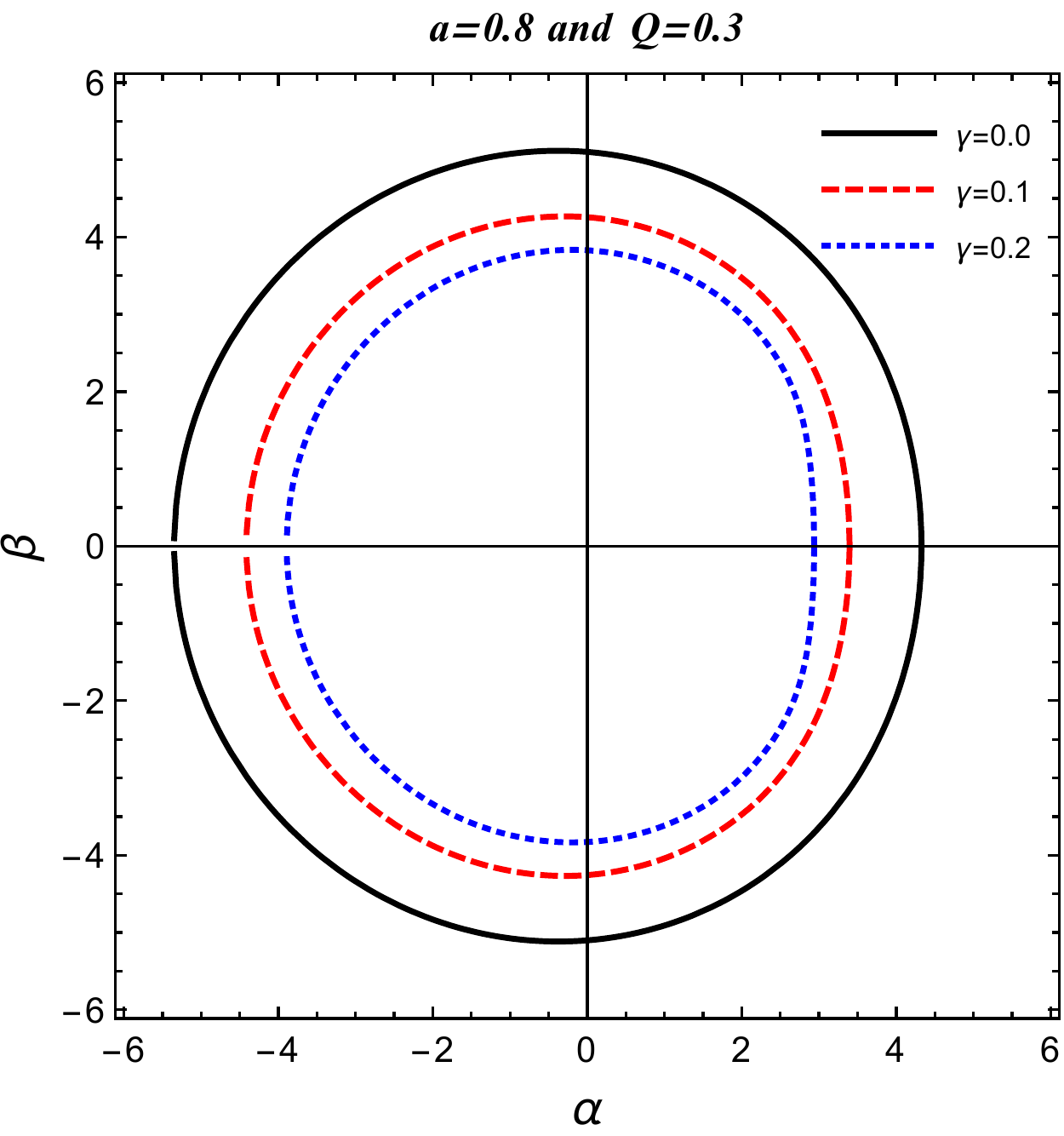}

       \includegraphics[scale=0.45]{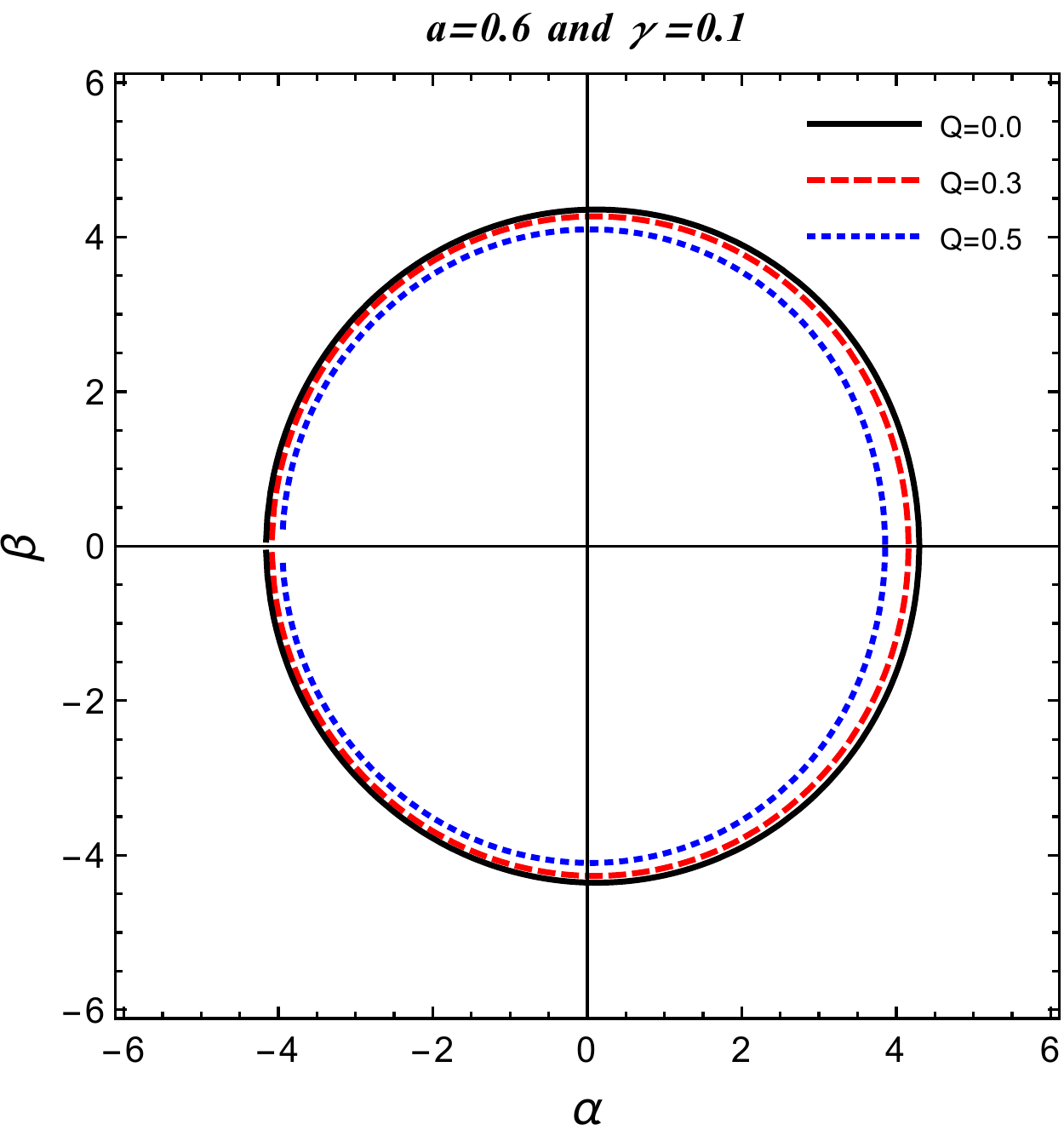}
   \includegraphics[scale=0.45]{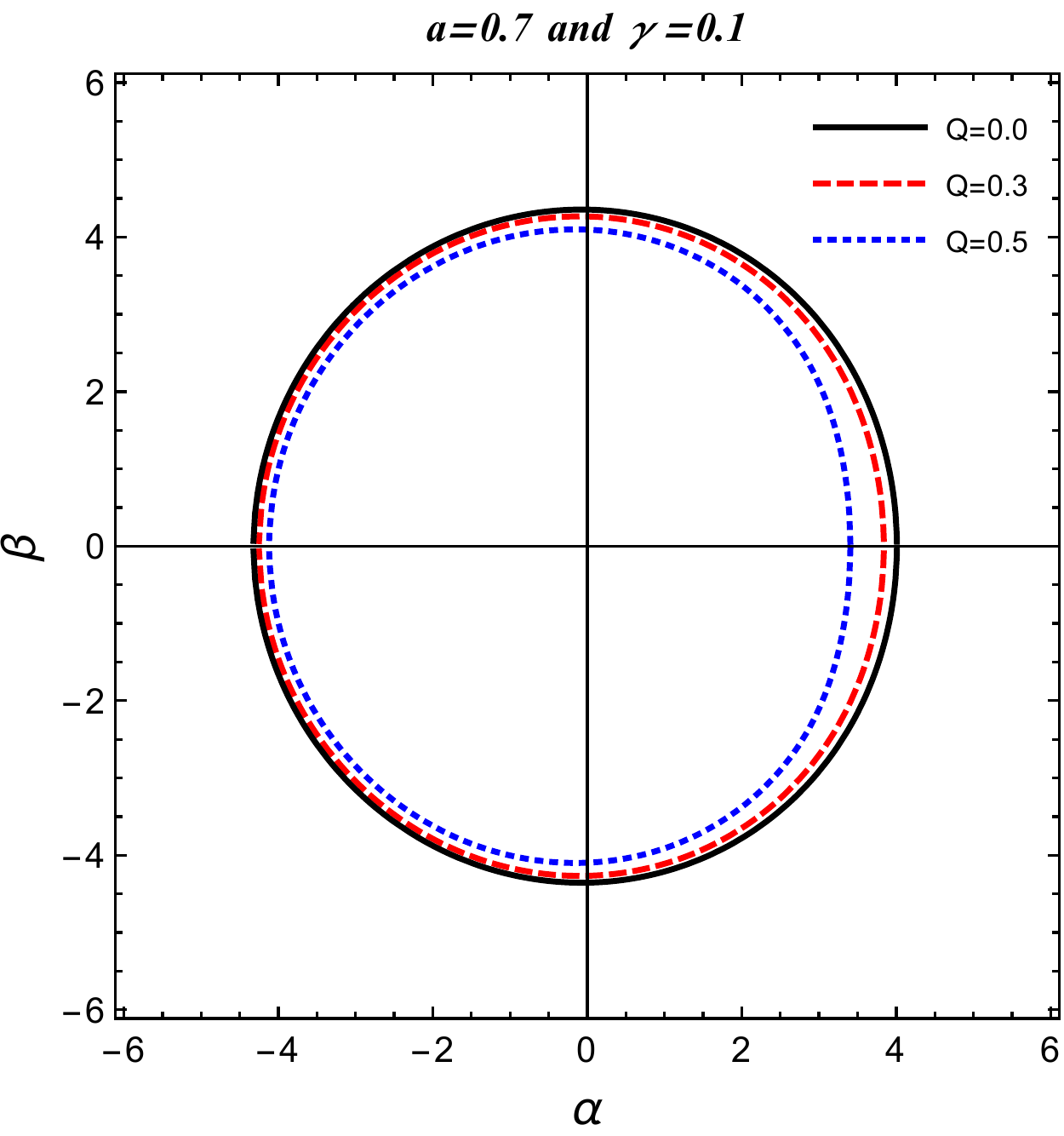}
    \includegraphics[scale=0.45]{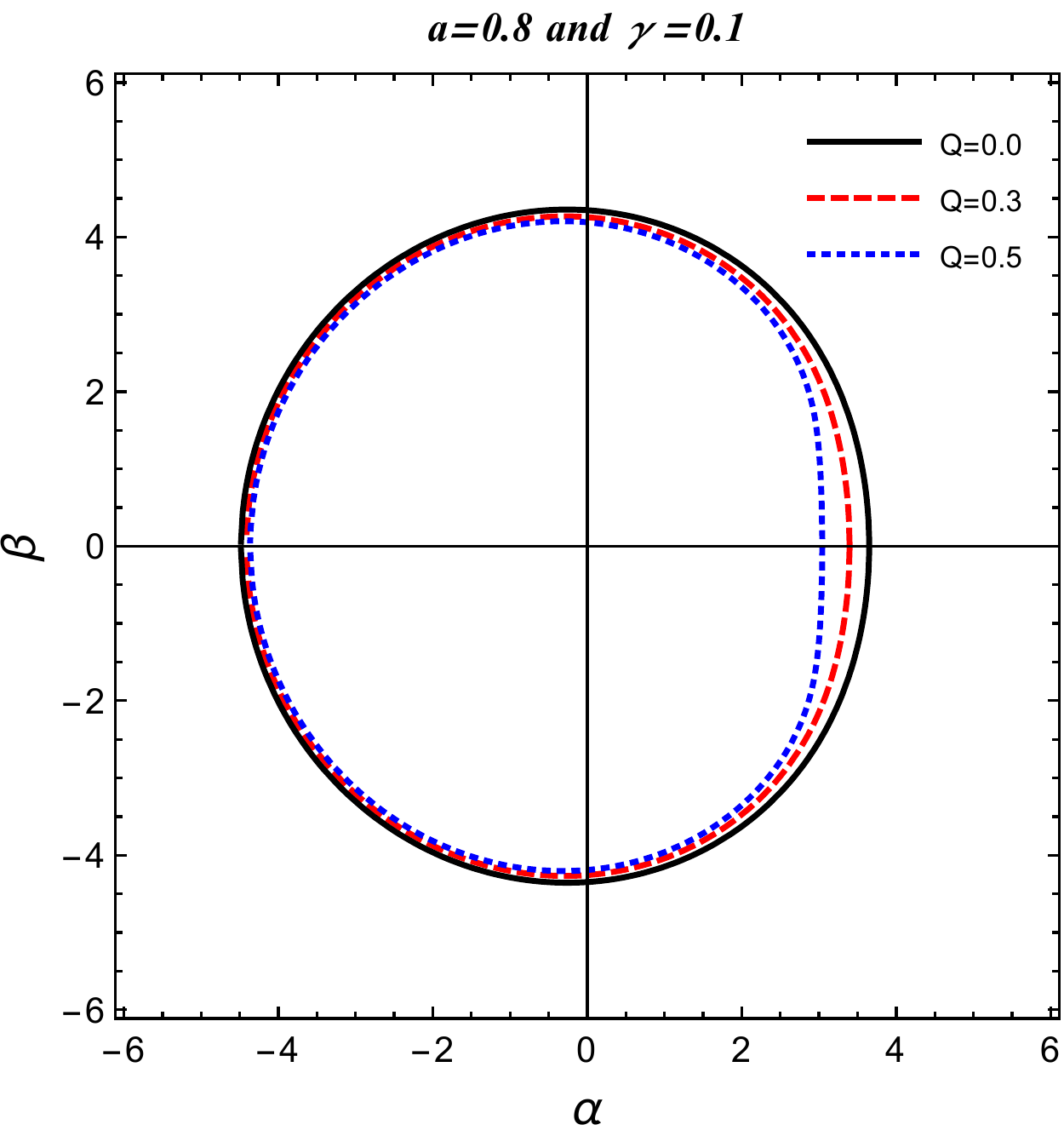}

    \includegraphics[scale=0.45]{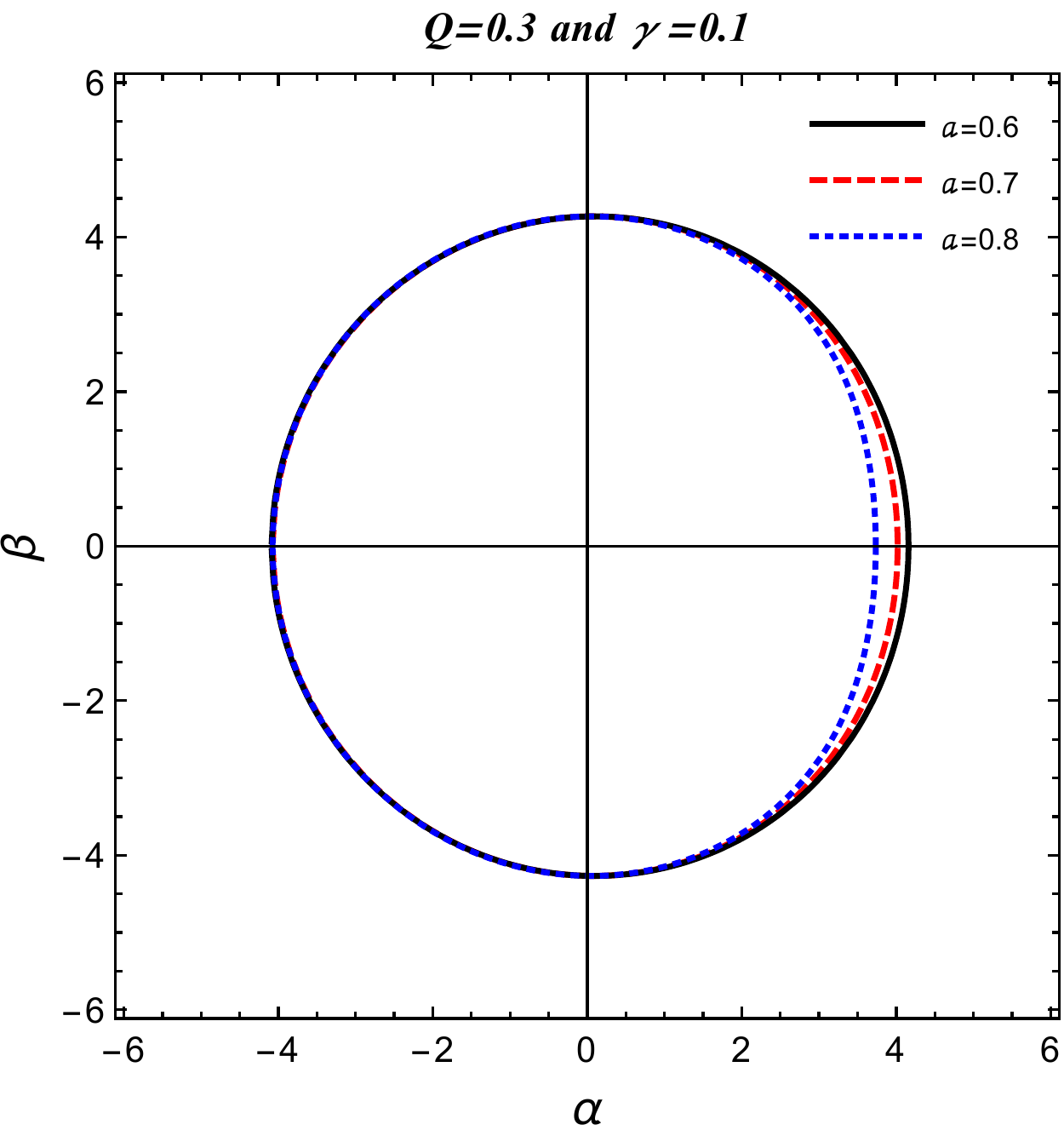}
   \includegraphics[scale=0.45]{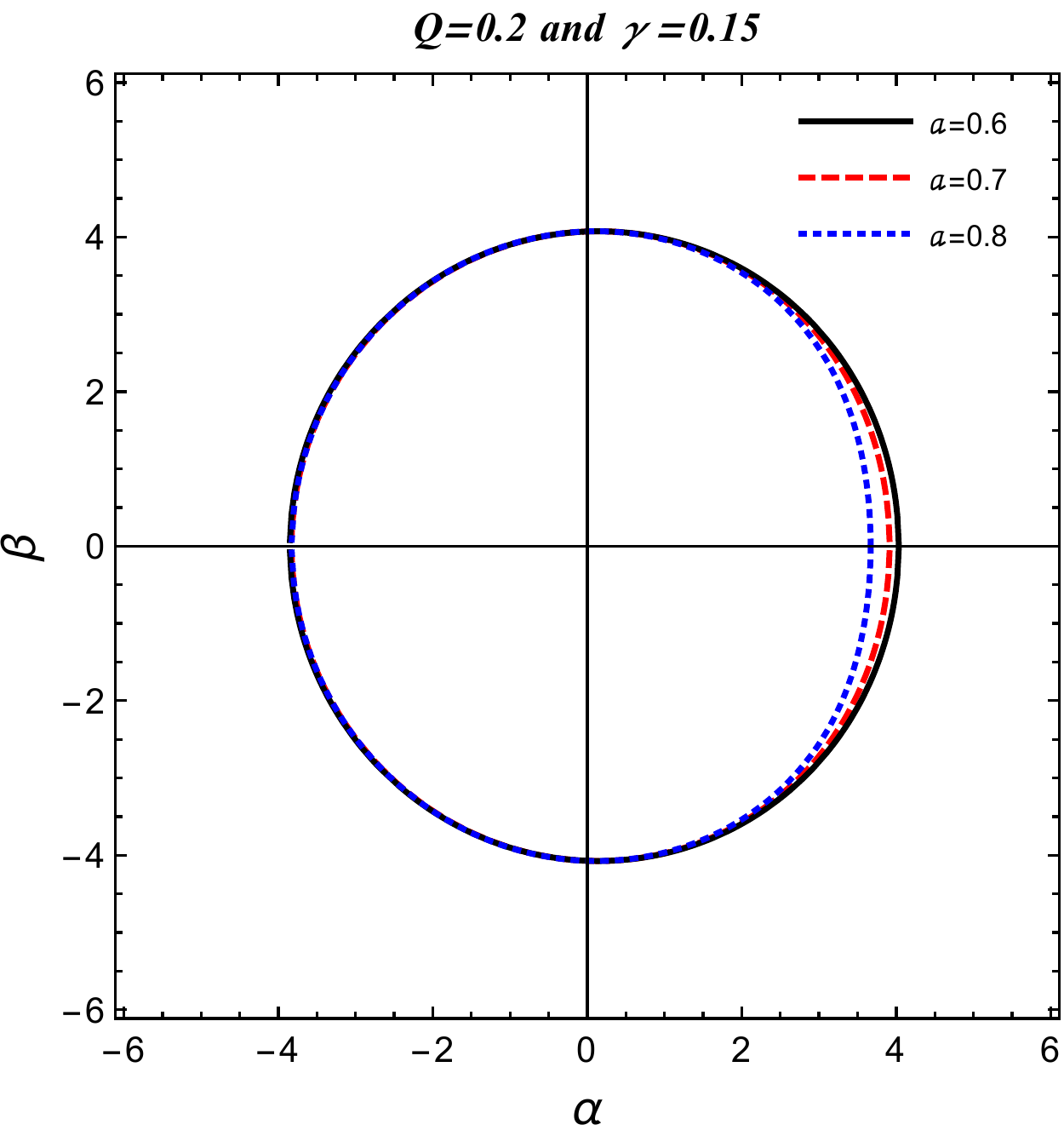}
    \includegraphics[scale=0.45]{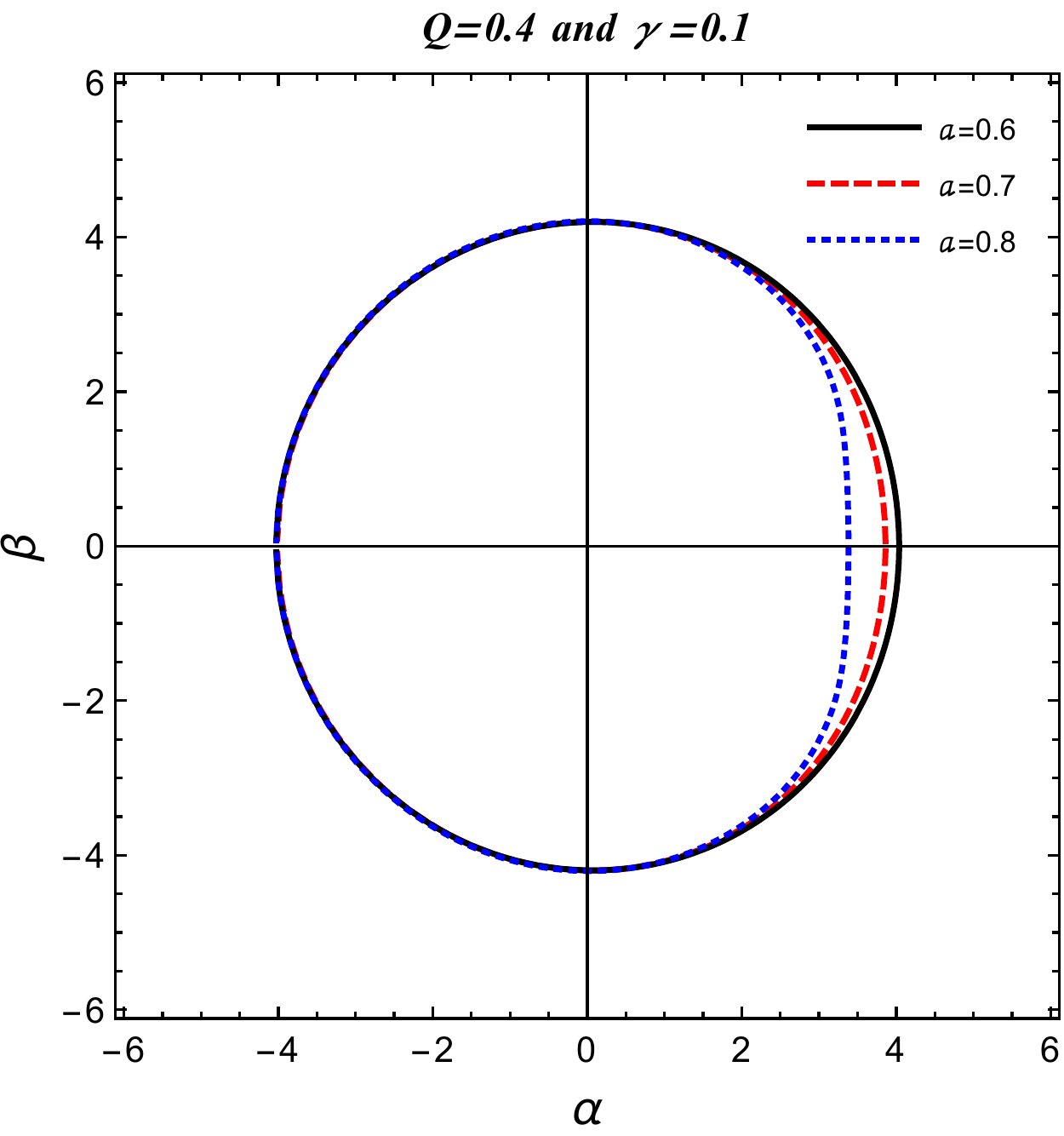}
    \end{center}
\caption{Shadow cast by charged rotating BH at different values of PFDM parameter $\gamma$, charge $Q$ and rotation parameter $a$.}\label{shadowrotating}
\end{figure*}

\section{Shadow of the BH}\label{Shadowsec}
In this section, we want to analyze the shadow of BHs for an observer at $r=\infty$, so we introduce the celestial coordinate $\alpha$ and $\beta$, which are \cite{Vazquez04a,Abdujabbarov13a}

\begin{eqnarray}\label{alpha}
\alpha=\lim_{{r}_{0}\rightarrow \infty} -{r}_{0}^{2}\left(\sin\theta_{0}\frac{d\phi}{dr}\right),
\end{eqnarray}
and
\begin{equation}  \label{beta}
\beta=\lim_{{r}_{0}\rightarrow \infty}{r}_{0}^{2}\frac{d\theta_0}{dr},
\end{equation}
where ${r}_{0}$ is
the distance from the BH to observer, the coordinate
$\alpha$ is the apparent perpendicular distance between the image and the axis of symmetry, and the coordinate $\beta$ is the apparent
perpendicular distance between
the image and its projection on the equatorial plane.

Now considering  that the observer is far away from the BH
and hence,  $r_0$ tends to infinity, and the angular
coordinate of the distant observer $\theta_0$ is the inclination
angle between the line of sight of the distant observer and the
axis of rotation of the central gravitating object. To
investigate the shadow two coordinates $(\alpha,
\beta)$ are introduced. Using Eq. (\ref{EOM}), in Eqs. (\ref{alpha}) and (\ref{beta}), and taking the limit of a far away observer, we found the celestial coordinates, as a function of the constant of motion, take the form
\begin{eqnarray}\label{AB}
\alpha &=& -\left( \xi \frac{1}{\sin\theta_0} \right),\nonumber \\\beta &=& \pm \sqrt{\eta +a^2 \cos^2\theta_0 - \xi^2 \cot^2\theta_0}.
\end{eqnarray}
In the equatorial plane $\theta_0=\pi/2$, then the celestial coordinates can be written as \begin{eqnarray}
\alpha &=& -\xi, \nonumber \\ \beta &=& \pm \sqrt{\eta}.\label{alphabetalast}
\end{eqnarray}

\begin{figure}
   \includegraphics[scale=0.6]{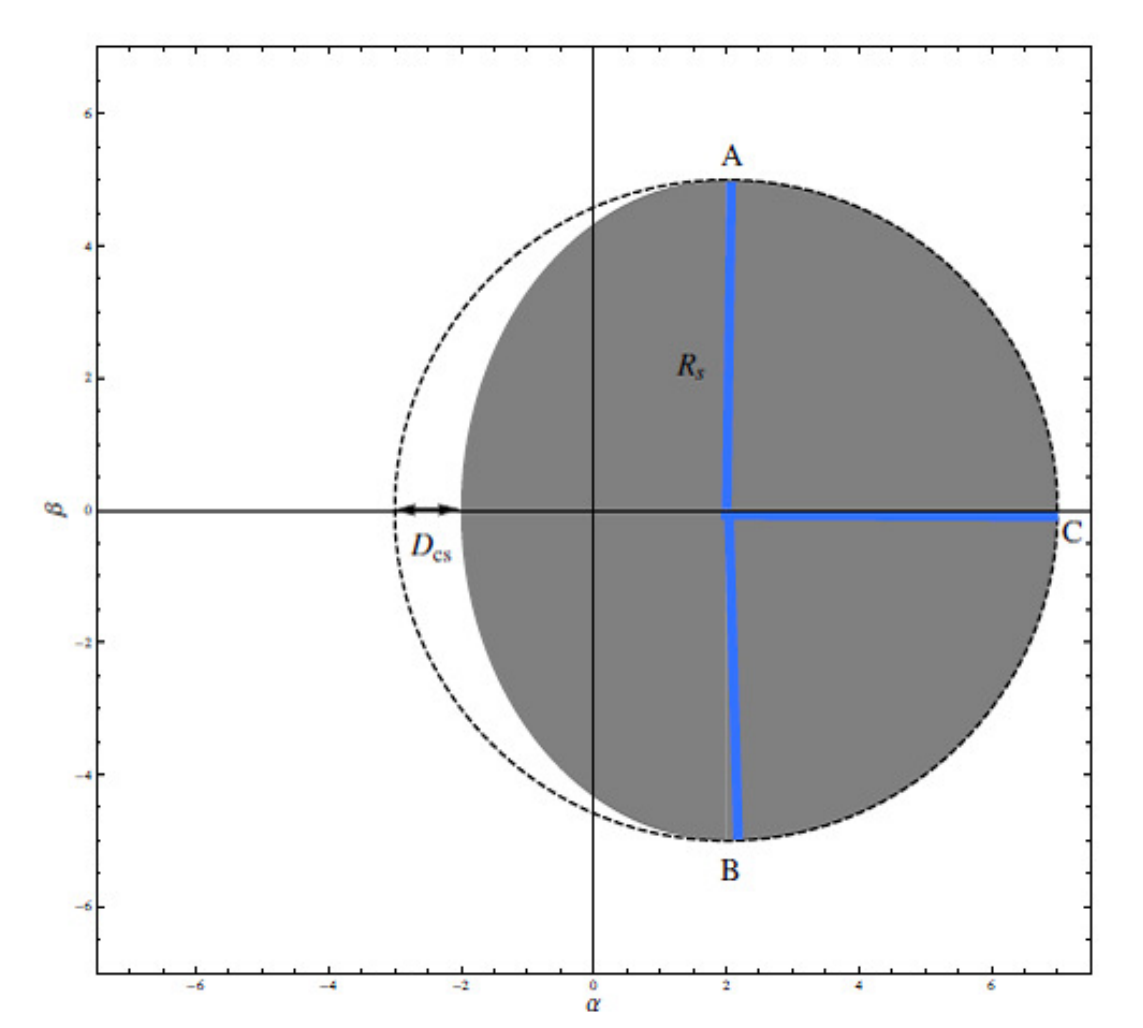}
\caption{The observables the radius $R_s$,
and the distortion parameter $\delta_s$ as described in the
Ref. \cite{Amarilla10a}.}\label{radi}
\end{figure}
\begin{figure*}
 \begin{center}
   \includegraphics[scale=0.5]{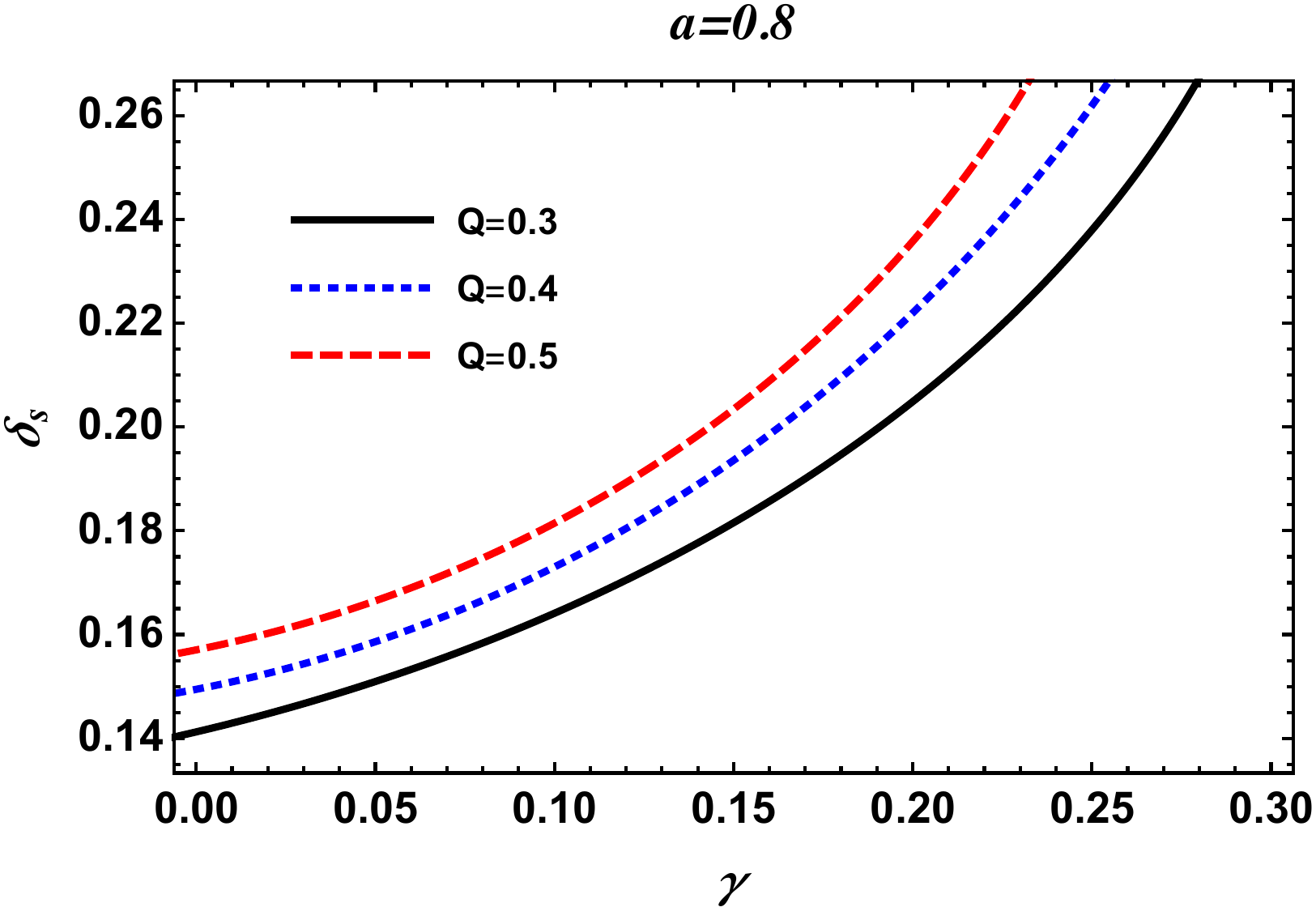}
   \includegraphics[scale=0.5]{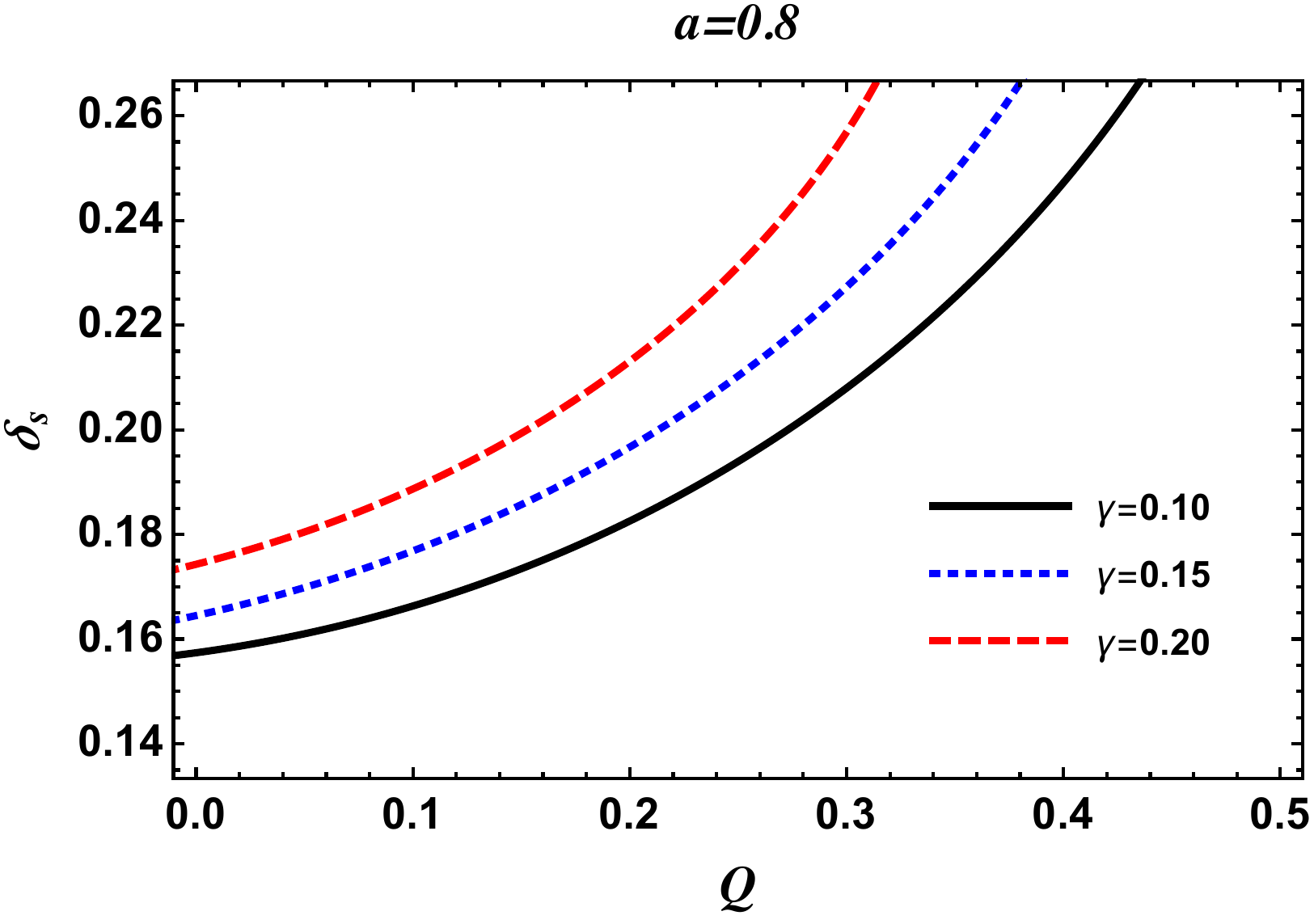}
  \end{center}
\caption{Plot showing the variation of distortion parameter $\delta_s$ with PFDM parameter $\gamma$ (left) and with charge $Q$ (right) for different value of $Q$ and $\gamma$ respectively.}\label{distortion}
\end{figure*}
\subsection{Non-rotating case($a=0$)}
To understand the effect of PFDM parameter we have studied here the non rotating charged BH case to compare with the rotating charged one. Using the Eq. (\ref{alphabetalast}), one can easily obtain the radius for the case of a non-rotating charged BH in the presence of PFDM, which reads as \cite{Atamurotov16a}
\begin{widetext}
\begin{eqnarray}
R_{sh}=
\frac{1}{\Big(\gamma -2 M+\gamma  \ln \Big(\frac{r}{\gamma }\Big)+2 r\Big)} \Big(2 r \Big(-12 M^2 r+8 M Q^2+4 \gamma  M r-4 \gamma  Q^2+4 r^3+\gamma ^2 r\Big) \nonumber \\ -4 \gamma  \ln \Big(\frac{r}{\gamma }\Big) \Big(-6 M r^2+2 Q^2 r+\gamma  r^2\Big)-6 \gamma ^2 r^2 \ln ^2\Big(\frac{r}{\gamma }\Big)\Big)^{1/2}
\end{eqnarray}
\end{widetext}
 To see the effect of charge and PFDM we have plotted the radius with respect to charge $Q$ and PFDM parameter $\gamma$ for varying $\gamma$ and $Q$ respectively in Fig.~\ref{shadowa0}. It is observed that the size decreases with increasing both $Q$ and $\gamma$, but the rate of decrease in radius is greater for PFDM parameter than that of charge.

\subsection{Rotating case}
For the visualization of the shape of the shadow we have plotted $\beta$ Vs $\alpha$ in Fig. \ref{shadowrotating}. The variation of the shape of the shadow for varying PFDM parameter $\gamma$ at fixed charge $Q$ and rotation parameter $a$, for varying charge $Q$ at fixed  $\gamma$ and $a$ and for different value of rotation parameter at fixed $\gamma$ and $Q$ is shown. It can be noted from the plots that the size of the shadow decreases with the increase in PFDM parameter at fixed $a$ and $Q$ with a distortion in the shape with changing both $a$ and $\gamma$, similarly for varying charge $Q$ also size of the shadow is decreasing along with distortion in the shape, but the rate of decrease in the size of shadow is lesser than that in the case of PFDM parameter. However, for varying rotation parameter at fixed $Q$ and $\gamma$, there is only distortion in the shape of the shadow.

To characterize the shape of the shadow of charged rotating BH in presence of PFDM we introduce two  observables the radius $R_s$ and distortion parameter $\delta_s$. Considering the shape of the shadow of the BH is a circle as shown in Fig. \ref{radi} (See reference \cite{Amarilla10a}). Let us consider that shadow as  a circle is  passing through $(A)$ (top), $(B)$ (bottom), and the most left end $C$ of its boundary. The coordinates to these points be $(\alpha_t, \beta_t)$, $(\alpha_b, \beta_b)$ and $(\alpha_r, 0)$ respectively. The radius $R_s$ of the circle defines the radius of shadow. The point $C$ on the circle represents the unstable retrograde circular orbit for
an observer on the equatorial plane. The distortion parameter
$\delta_s$ is defined as
\begin{equation}
\delta_{s}=\frac{D_{cs}}{R_s},
\end{equation}
here $D_{cs}$ is the difference between the right endpoints of the shadow. The observable $R_s$ is defined as
\begin{equation}
R_s = \frac{(\alpha_t - \alpha_r)^2 + \beta_t^2}{2(\alpha_t - \alpha_r)}.
\end{equation}

In Fig.~\ref{distortion}, the observable $\delta_s$  is
plotted as a function of the PFDM parameter $\gamma$ and charge $Q$. We
observe that the observable $\delta_s$ gives distortion in the shape of the shadow with increasing PFDM parameter and charge both.

\section{Emission energy}\label{emissionenergysec}
\begin{figure*}
 \begin{center}
   \includegraphics[scale=0.6]{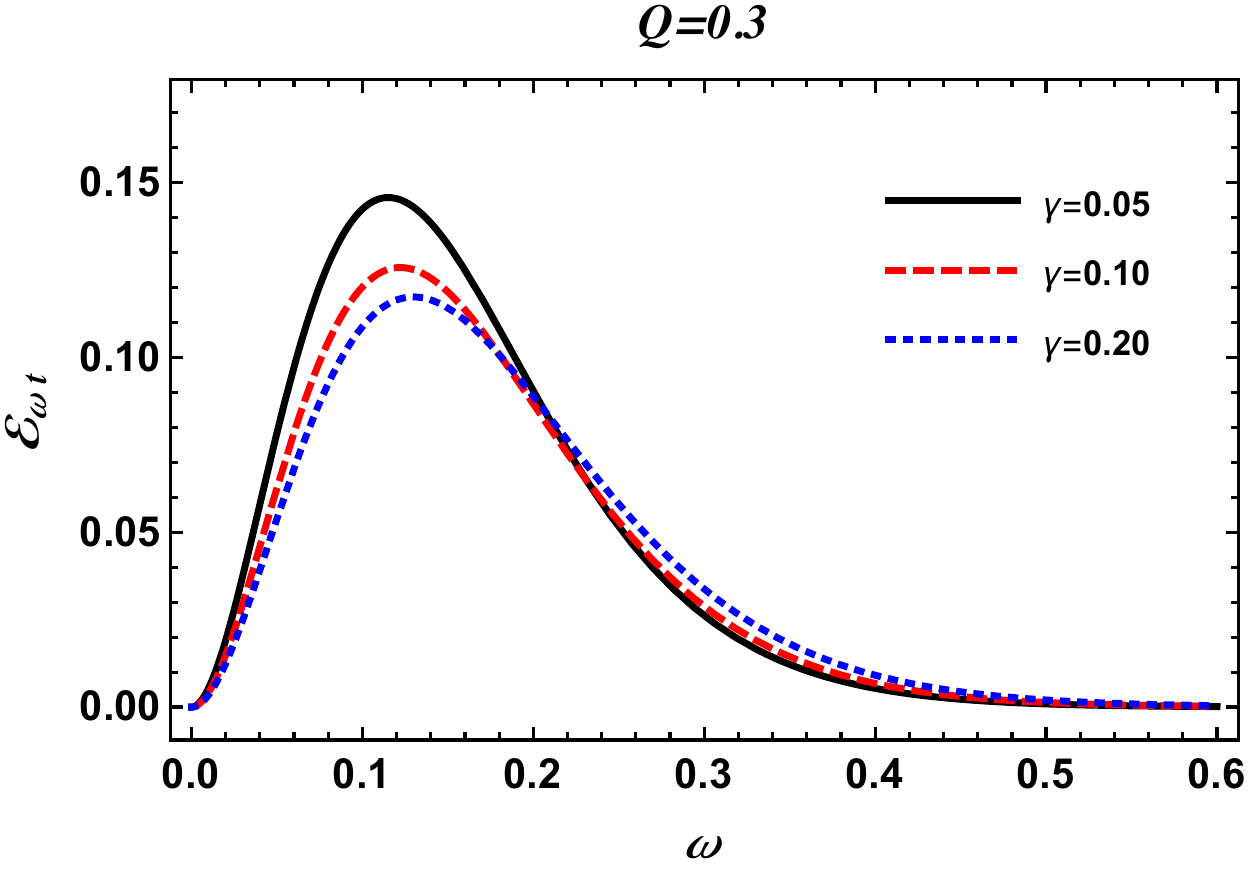}
   \includegraphics[scale=0.6]{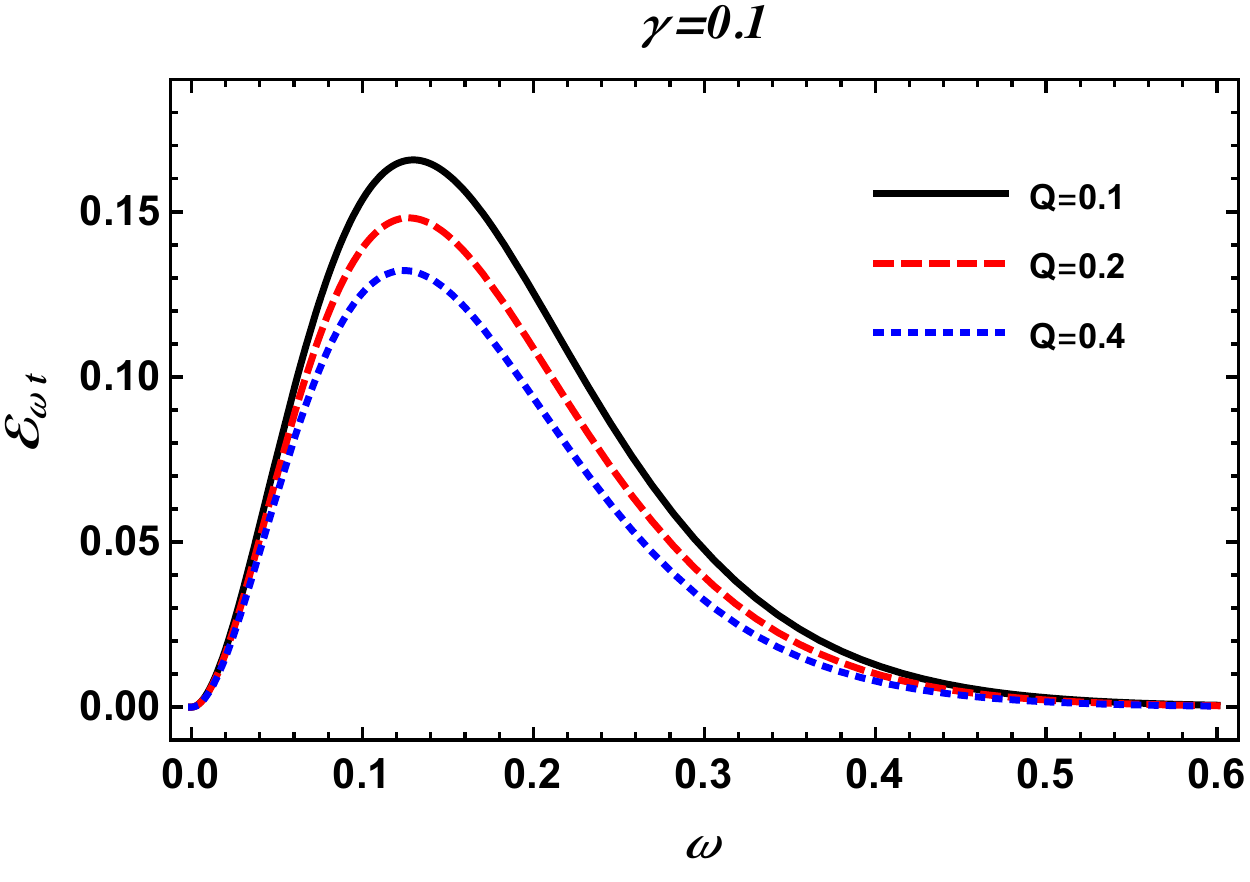}
    \end{center}
\caption{Plots showing the rate of energy emission varying with
the frequency for different values of the charge $Q$ (Left panel) and  PFDM parameter $\gamma$ (Right panel). }\label{energy}
\end{figure*}
It is well known that BHs emit thermal radiations which
lead to a slow decrease in mass of the BH until
it completely annihilates, Here, we compute the energy emission rate of rotating charged BH in presence of PFDM using the relation \cite{Wei13a}
\begin{equation}
\frac{d^2 \cal E(\omega)}{d\omega dt}= \frac{2 \pi^2 \sigma_{lim}}{\exp{\omega/T}-1}\omega^3,
\end{equation}where $T=\kappa/2 pi$ is the Hawking temperature and $\kappa$ is the surface gravity.  $\sigma_{lim}$ is the limiting constant value for a spherically symmetric BH around which the absorption cross section oscillates. The limiting constant is expressed as \cite{Wei13a}
\begin{equation}
\sigma_{lim} \approx \pi R_s^2. \nonumber
\end{equation}Hence
\begin{eqnarray}
\frac{d^2 \cal E(\omega)}{d\omega dt}=\frac{2\pi^3 R_s^2}{e^{\omega/T}-1}\omega^3. \nonumber
\end{eqnarray}
The variation of energy emission with PFDM parameter and charge is visualises in Fig.~\ref{energy} and it is seen that with the increase in the values of PFDM parameter and BH charge $Q$, the peak of energy emission rate ( where \begin{equation}{\cal E}_{\omega t}=\frac{d^2 {\cal E}(\omega)}{d\omega dt} \nonumber \end{equation}  is introduced) decreases.

\section{Deflection angle of relativistic massive particles}\label{deflectionsec}

In this section we will use the slowly rotating charged BH metric surrounded perfect fluid to compute the deflection of relativistic neutral particles and light. Toward this purpose, we shall follow the recent work of Crisnejo, Gallo and Jusufi \cite{Crisnejo:2019ril}.  We can use a correspondence between the motion photons in plasma and the motion of a test massive particle in the same background find the deflection angle. Having a stationary and axisymmetric spacetime $(\mathcal{M},g_{\alpha\beta})$, the non-magnetized plasma is characterized with the refractive index $n$ which can be written as
\begin{equation}\label{refra-index}
    n^2(x,\omega(x))=1-\frac{\omega_e^2(x)}{\omega^2(x)},
\end{equation}
where $\omega(x)$ gives the photon frequency measured by an observer following a timelike Killing vector field and $\omega_e(x)$ is known as the plasma frequency. It is worth noting that the quantity $\omega(x)$ in terms of the gravitational redshift is given by
\begin{equation}
    \omega(x)=\frac{\omega_\infty}{\sqrt{-g_{00}}}.
\end{equation}

According to this correspondence, the electron frequency of the plasma  $\hbar\omega_e$ can be linked to the rest mass $m$ of the massive particle and the total energy $E=\hbar \omega_\infty$ of a photon with the total energy $E_\infty=m/(1-v^2)^{1/2}$. In what follows we are going to use this idea and apply the Gauss-Bonnet theorem to compute the deflection angle.\\

Let $D\subset S$ be a regular domain of an oriented two-dimensional surface $S$ with a Riemannian metric $\tilde{g}_{ij}$, the boundary of which is formed by a closed, simple, piece-wise, regular, and positive oriented curve $\partial D: \mathcal{R}\supset I\to D$. Then \cite{Crisnejo:2019ril},
\begin{equation}
    \int\int_D \mathcal{K}dS+\int_{\partial D} k_g dl +\sum_i \epsilon_i = 2\pi\chi(D), \ \ \ \sigma\in I.
\end{equation}

In the last equation, $\chi(D)$ is known as the Euler characteristic number (in our case it will be one), $\mathcal{K}$ is known as the Gaussian
curvature over the optical domain, $k_g$ is the geodesic curvature, and $\epsilon_i$ represents the corresponding exterior angle in the i-th vertex. We can now formulate the above theorem using the deflection angle $\hat{\alpha}$, as follows \cite{Crisnejo:2019ril}
\begin{equation}
    \int_0^{\pi+\hat{\alpha}}\bigg[\kappa_g \frac{d\sigma}{d\phi}\bigg]\bigg|_{C_R}d\phi=\pi -\bigg(\int\int_{D_R}\mathcal{K}dS + \int_{\gamma_p} k_g dl \bigg),
\end{equation}
and we need to take the limit $R\to\infty$. Since our metric is asymptotically flat,  the following condition holds
$[k_g\frac{d\sigma}{d\phi}]_{C_R}\to 1$ if the radius of $C_R$ tends to infinity. Finally, we can find a simple form to express the deflection angle which reads
\begin{equation}\label{alpha-general}
    \hat{\alpha}=-\int\int_{D_R}\mathcal{K}dS - \int_{\gamma_p} k_g dl.
\end{equation}
Basically we can use the last expression to evaluate the deflection angle of particles.  Due to the spherical symmetry of the problem, we will study the problem in the equatorial plane, hence we set $\theta=\pi/2$. Now we need to compute  the geodesic curvature which can be calculate by the expression
\begin{equation}\label{eq:kgasada}
    k_g =-\frac{1}{\sqrt{\hat{g} \hat{g}^{\theta\theta} }} \partial_r\hat{\beta}_\phi,
\end{equation}
in which $\hat{g}$ is the determinant of $\hat{g}_{ab}$. The second quantity that we need and has to be calculated, is the Gaussian optical curvature which can be found by the Ricci scalar of the optical metric using a simple formula $\mathcal{K}=R/2$. In general, the metric on the equatorial plane has the form
\begin{equation}\label{eq:metricoriginal}
\begin{split}
    ds^2&=-A\,dt^2+B\,dr^2-2Hdtd\phi +D d\phi^2.
    \end{split}
\end{equation}
From this form, we can find the corresponding Finsler-Randers metric which has the form \cite{Crisnejo:2019ril}
\begin{equation}
    \mathcal{F}(x,\dot{x})=\sqrt{\hat{g}_{ab} \dot{x}^a \dot{x}^b}+ \bm{\hat\beta}_a \dot{x}^{a},
\end{equation}
with
\begin{eqnarray}
    \hat{g}_{ab} dx^a dx^b&=&n^2[(\frac{B}{A})dr^2\nonumber+\frac{AD+H^2}{A^2}d\phi^2],\nonumber\\
    \bm{\hat\beta}&=&-\frac{H}{A}d\phi.
\end{eqnarray}
We shall focus only on linear order of $a$, hence for the metric components it follows
\begin{eqnarray}
    A(r)&=& 1-\frac{2 M}{r} +\frac{Q^2}{r^2} + \frac{\gamma}{r} \ln(\frac{r}{\gamma}),\\
    B(r)&=& \frac{1}{1-\frac{2 M}{r} +\frac{Q^2}{r^2} + \frac{\gamma}{r} \ln(\frac{r}{\gamma})},\\
    D(r)&=& r^2,\\
    H(r)&=& \frac{2 a M}{r}-\frac{a \gamma}{r}\ln (\frac{r}{\gamma}),
\end{eqnarray}
along with the refractive index  which can be written as
\begin{eqnarray}
    n^2(r)&=& 1-(1-v^2)A(r).
\end{eqnarray}

For the deflection angle it follows that
\begin{equation}\label{alpha-pm}
    \hat{\alpha}_{\text{mp}}=-\iint_{D_r}\mathcal{K}dS-\int_{R}^S k_g dl,
\end{equation}
where $l$ is an affine parameter (see for details \cite{Crisnejo:2019ril}). Moreover $S$ stands for the source and $R$ for the receiver.  The slowly rotating PFDM BH metric in the equatorial plane has can be written as
\begin{eqnarray}\notag
    ds^2 &=&-( 1-\frac{2 M}{r} +\frac{Q^2}{r^2} + \frac{\gamma}{r} \ln(\frac{r}{\gamma}))dt^2-2a \left(\frac{2M}{r}\right) dt d\phi \\
    &+& \frac{dr^2}{( 1-\frac{2 M}{r} +\frac{Q^2}{r^2} + \frac{\gamma}{r} \ln(\frac{r}{\gamma}))} + r^2d\phi^2.
\end{eqnarray}
Note that we have neglected the term $a \gamma$ to simplify further the problem. The Gaussian optical curvature is found to be
\begin{eqnarray}
    \mathcal {K} & \simeq & -\frac{M}{r^3 v^2 }(1+\frac{1}{v^2})-\frac{Q^2}{r^4 v^2 }(1+\frac{2}{v^2}) \\
    &-& \frac{\gamma}{2 v^2 r^3}(1+\frac{2}{v^2})+\frac{\ln({\frac{r}{\gamma}})}{2 v^2 r^3 }(1+\frac{1}{v^2})
\end{eqnarray}
For the geodesic curvature it can be shown that
\begin{equation}
    \Big[k_gdl\Big]_{r_{\gamma}} \simeq -\frac{2 a M}{v b^2}\sin\phi d\phi.
\end{equation}
One can compute the deflection angle by the following equation
\begin{equation}\label{alpha-pm}
   \hat{\alpha}_{\text{mp}}=-\int_{0}^{\pi}\int_{b/\sin(\phi)}^{\infty}\mathcal{K} \sqrt{\det \hat{g}} dr d\phi-\int_{0}^{\pi}s\frac{2 a M}{v b^2}\sin\phi d\phi,
\end{equation}
in the last equation  $s=+1$ stands for the prograde orbits and $s=-1$ for retrograde orbits. Finally with the help of \eqref{alpha-pm} for the deflection angle of particles we obtain
\begin{eqnarray}\label{eq:eRN}\notag
    \hat{\alpha}_{\text{mp}} &=& \frac{2M}{b}\bigg(1+\frac{1}{v^2}\bigg)-\frac{\pi Q^2}{4 b^2}\bigg(1+\frac{2}{v^2}\bigg)\\\notag
    &+& \frac{\gamma}{v^2 b}\Big[\ln(2)(1+v^2)-\ln(\frac{b}{\gamma})(1+v^2)-v^2\Big]\\
    &-& \frac{4saM}{b^2 v}.
\end{eqnarray}
The effect of electric charge clearly decreases the deflection angle, while the effect of the PFDM is more complicated since it depends not only on the PFDM parameter $\gamma$, but also on the ratio between the impact parameter and $\gamma$ as well as the velocity of the particle $v$.

\section{Effect of perfect fluid dark matter on Einstein rings}\label{ringsec}
 If we consider the special case $v=1$ in Eq. (50), we obtain the deflection angle of light in a PFDM reported in \cite{Haroon19a}. In order to simplify further the notation let us introduce the following equation
 \begin{equation}
     \frac{b}{\gamma}=10^n,
 \end{equation}
 where $n\in \mathrm{R}$. It follows that
\begin{equation}
    \hat{\alpha}_{\text{light}} =\frac{4M}{b}-\frac{3 \pi Q^2}{4 b^2}+ \frac{\gamma}{ b}[2 \ln(2)-4.6\, n-1]- \frac{4saM}{b^2},
\end{equation}
which generalizes the result obtained in Ref. \cite{Haroon19a}. In leading order terms we see that the dominating terms are the term with the BH mass and the dark matter term. In addition, we see that when $n$ increases the deflection angle of light gets smaller.  By taking the lens equation given as \cite{Bozza:2008ev}
\begin{equation}\label{ImagePositions}
    D_{OS}\tan\beta=\frac{D_{OL}\sin\theta-D_{LS}\sin(\hat{\alpha}-\theta)}{\cos(\hat{\alpha}-\theta)}.
\end{equation}
One can use this equation to find the positions of the weak field images. Since we are interested in the weak deflection approximation, it follows that
\begin{eqnarray}\label{EinsteinRing}
    \beta=\theta-\frac{D_{LS}}{D_{OS}}\hat{\alpha}.
\end{eqnarray}
The Einstein ring is formed if $\beta=0$. Now using the fact that $\hat{\alpha}\ll1 $, we obtain the angular radius of the Einstein ring
\begin{eqnarray}\label{EinsteinRing1}
    \theta_{E}\simeq\frac{D_{LS}}{D_{OS}}\hat{\alpha}(b).
\end{eqnarray}
By construction, it follows that $D_{OS}=D_{OL}+D_{LS}$, where we will take as an example the BH located at our galactic center Sgr A$^{*}$.  Furthermore we shall assume the following conditions $D_{LS}= D_{OL}/2$ hence $D_{OS}=3D_{OL}/2$.  To first order of approximation, if we can neglect the rotation  in the deflection angle and using the relation $b=D_{OL}\sin{\theta}\simeq D_{OL}\theta$, the bending angle in the hypothesis of small angles. For such case, the contribution of the charge and rotation will be very small and the main contribution as can be seen from Eq. (52) comes from the BH mass and the surrounding dark matter. The angular size in such a case gives
\begin{eqnarray}
    \theta_E=\sqrt{\frac{D_{LS}}{D_{OS}D_{OL}}\left(4M+\gamma [2 \ln(2)-4.6\, n-1]\right)}.
\end{eqnarray}

In Table I, we estimate the numerical value of the size of Einstein rings for different values of $\gamma$. It is shown that, by increasing the dark matter parameter $\gamma$, the size of the Einstein ring decreases. Compared to the case of Schwarzschild BH which has the size of Einstein ring $\theta_E^{Sch}=1.186098481$ arcsec, we conclude that in the presence of the perfect fluid dark matter the size of Einstein rings is smaller.

\begin{table}[tbp]
        \begin{tabular}{|l|l|l|l|l}
        \hline
   \multicolumn{1}{|c|}{ } &  \multicolumn{1}{c|}{  $\gamma=0.1$ }
    & \multicolumn{1}{c|}{  $\gamma=0.15$ } &   \multicolumn{1}{c|}{  $\gamma=0.2$ }  \\
    \hline
  $n$ & $\theta_E (\text{arcsec})$  & $\theta_E (\text{arcsec})$ & $\theta_E (\text{arcsec})$  \\
        \hline
1 &  1.1218870090  & 1.0883615610  & 1.0537700490   \\
2 &  1.0473037080  & 0.9704909999 & 0.8870515797   \\
3 & 0.9669848250  & 0.8361666543 & 0.6806538735  \\
4 & 0.8793601348 &  0.6756452962 & 0.3737898953   \\
        \hline
        \end{tabular}
         \caption{  Numerical values of the angular size of Einstein rings. We have used the central black hole with mass $M=4.31\times 10^{6}M_{\odot}$. The distance is taken $D_{OL}=8.33$ kpc from the Sgr A$^{*}$ to the observer. }
\end{table}

\section{Conclusions}\label{conclusionsec}
There are several observational parameters to study properties of BH but there is no exact data in which only charge or rotational parameters of BHs are discussed. Here we have tried to analyse theoretically the charged rotating BH in presence of PFDM and hence we have investigated the shape of the shadow for same to see the effect of PFDM parameter on the horizon, effective potential, shape of the shadow and energy emission. Further, we have also discussed about the deflection angle of relativistic massive particles and the effect of PFDM parameter on the Einstein rings for charged rotating BH.
 With this we obtained the following results:
\begin{itemize}
\item{We study the variation of horizon radius with rotation parameter and it is observed that both charge $Q$ and PFDM parameter $\gamma$ shows the influence on the horizon of the BH.}
\item{We have plotted effective potential with respect to radius in Fig. \ref{effpotential} and it is seen
that with the increase in the value of PFDM parameter $\gamma$, charge $Q$ and rotation parameter $a$ the peak of the graph is shifting towards the left, which signifies that
the peak value is increasing with the increase in the value of these parameters.}
\item{We have observed that the size of the shadow in case of non rotating BH decreases with the increase in value of PFDM parameter $\gamma$ as well as BHs charge $Q$.}
\item{The effect of $\gamma$ on the shape and size of shadow for charged rotating BH  shows that the size decreases with the increase in value of $\gamma$ and $Q$ along with a distortion in the shape of the shadow in comparison to the shadow for usual charged rotating case in GR.}
\item{The rate of distortion in the shape of the shadow has also been observe by plotting the observable $\delta_s$ with respect to $\gamma$ and $Q$ in Fig. \ref{distortion} and it can be seen clearly that distortion is increasing with the increase in both the parameters.}
\item{The dependence of the energy emission rate
on the frequency for the different values of charge and
PFDM parameter is investigated in Fig. \ref{energy}.
It is observed that the rate of emission is higher for the small
value of both $\gamma$ and $Q$. Thus, a large amount of energy is liberated at low value of $\gamma$ and $Q$.}
\item{We have found that the PFDM affects the deflection angle of massive particles as well as light rays. The PFDM has an effect on the deflection angle in leading order term. This effect is mostly encoded in the ratio between the impact parameter and the PFDM parameter $b/\gamma$. The bigger this ratio is, we found that the smaller the deflection angle becomes. It has been shown that the size of Einstein rings decreases with the increase of the PFDM parameter $\gamma$. Compared to the Schwarzschild BH, due to thee effect of PFDM the size of Einstein radius is significantly smaller. The effect of charge and rotation are negligible.}
\end{itemize}
All the above mentioned results obtained are compared to the case of usual Kerr-Newman BH ($\gamma \to 0$), Kerr BH ($\gamma \to 0$ and $Q \to 0$) and Schwarzschild BH ($\gamma \to 0$, $Q \to 0$ and $a \to 0$) in GR in the prescribed limit.

\section*{Acknowledgements}
The authors would like to thank the referees for there valuable comments which help us to improve the paper. FA acknowledges the support of INHA University in Tashkent and this research is partly supported by the Research Grant F-FA-2021-510 of the Uzbekistan Ministry for Innovative Development. UP would like to thank University Grant Commission (UGC), New Delhi for DSKPDF through Grant No. F.4-2/2006(BSR)/PH/18-19/0009. and UP would also like to acknowledge the facilities at ICARD, Gurukula Kangri (Deemed to be University), Haridwar, India.

{}

\end{document}